\documentclass[12pt]{article}
\usepackage{graphicx}
\usepackage{amsfonts}
\usepackage{amssymb,amsmath}
\usepackage{color}

\newcommand{\nn}{\nonumber \\}
\newcommand{\bea}{\begin{eqnarray}}
\newcommand{\eea}{\end{eqnarray}}

\makeatletter
\@addtoreset{equation}{section}

\makeatother
\usepackage{multicol}
\usepackage{color}
\definecolor{rosso}{cmyk}{0,1,1,0.3}
\definecolor{verde}{cmyk}{0.8,0,0.6,0.25}
\definecolor{bluc}{cmyk}{1,0.4,0,0.1}
\definecolor{blucc}{cmyk}{0.8,0.3,0,0}

\def\be{\begin{equation}}
\def\ee{\end{equation}}
\def\fr{\frac}
\def\der{\partial}
\def\tr{{\rm tr}}
\def\({\left(}
\def\){\right)}
\def\1{^{(1)}}
\def\2{^{(2)}}

\def\<{\langle}
\def\>{\rangle}
\newcommand{\Slash}[1]{{\ooalign{\hfil/\hfil\crcr$#1$}}}

\begin{document}

\begin{titlepage}

\begin{flushright}
UT-12-28\\
\end{flushright}

\vskip 1.35cm

\begin{center}

{\large \bf  
Dynamics of oscillating scalar field in thermal environment
}

\vskip .45in

{
Kyohei Mukaida$^{(a)}$
and
Kazunori Nakayama$^{(a,b)}$
}

\vskip .45in

{\em
$^a$Department of Physics, University of Tokyo, Bunkyo-ku, Tokyo 113-0033, Japan \vspace{0.2cm}\\
$^b$Kavli Institute for the Physics and Mathematics of the Universe,
University of Tokyo, Kashiwa 277-8583, Japan \\
}

\end{center}

\vskip .4in

\begin{abstract}

There often appear coherently oscillating scalar fields in particle physics motivated cosmological scenarios,
which may have rich phenomenological consequences.
Scalar fields should somehow interact with background thermal bath in order to decay into radiation at an appropriate epoch,
but introducing some couplings to the scalar field makes the dynamics complicated.
We investigate in detail the dynamics of a coherently oscillating scalar field,
which has renormalizable couplings to another field interacting with thermal background.
The scalar field dynamics and its resultant abundance are significantly modified
by taking account of following effects:
(1) thermal correction to the effective potential,
(2) dissipation effect on the scalar field in thermal bath,
(3) non-perturbative particle production events and
(4) formation of non-topological solitons.
There appear many time scales depending on the scalar mass, amplitude, couplings and
the background temperature, which make the efficiencies of these effects non-trivial.

\end{abstract}

\end{titlepage}

\setcounter{page}{1}

\tableofcontents

%%%%%%%%%%%%%%%%%%%%%%%%%%%%%
\section{Introduction} 
\label{sec:introduction}
%%%%%%%%%%%%%%%%%%%%%%%%%%%%%

Scalar fields play important roles in particle physics and cosmology.
The standard model Higgs boson, which has been recently discovered, develops 
a vacuum expectation value (VEV) and spontaneously breaks gauge symmetry.
In the supersymmetric (SUSY) standard model (SM), there are many scalar fields
as superpartners of SM quarks and leptons.
These scalar fields may have large VEVs in the early Universe
and their dynamics may have significant effects on the baryon asymmetry of the Universe
through the Affleck-Dine mechanism~\cite{Affleck:1984fy}.

Inflation in the very early Universe is considered to be caused by a scalar field, called inflaton,
which slowly rolls down the scalar potential~\cite{Lyth:2009zz}.
After inflation, the inflaton decays into radiation and then hot thermal Universe begins.
The inflaton may also be responsible for the primordial density perturbation which seeds the rich structure of the
present Universe.
Instead, another scalar field, called curvaton, can explain the primordial density perturbation~\cite{Mollerach:1989hu,Linde:1996gt,Lyth:2001nq,Moroi:2001ct}.
The curvaton also decays into radiation and fluctuations in the curvaton sector are converted
to the adiabatic perturbation in the radiation.

In these scenarios, the understandings of the dynamics of coherently oscillating scalar fields
are essential for deriving their phenomenological consequences.
For example, the inflaton or curvaton must dissipate its energy into the radiation in order for 
a hot thermal Universe to begin well before the big-bang nucleosynthesis. 
It inevitably requires interactions between the inflaton/curvaton and other fields, which somehow will be thermalized.
A conventional picture of this {\it reheating} process is that the oscillating scalar field decays into lighter particles
and the produced particles are eventually thermalized by gauge and/or Yukawa interactions.
This simple picture, however, does not necessarily hold in general.
Depending on the couplings between the scalar ($\phi$) and other fields ($\chi$),
many non-trivial issues appear which makes this topic far more complicated.

Let us assume the simplest Yukawa coupling between the scalar $\phi$ and a fermion $\chi$, as
\begin{equation}
	\mathcal L = \lambda \phi \bar\chi_{\rm L} \chi_{\rm R} + {\rm h.c.},  \label{yukawa}
\end{equation}
in order to induce the $\phi$ decay into radiation, where $\lambda$ is a coupling constant and
$\chi$ is assumed to have gauge or Yukawa interactions with 
SM particles.\footnote{
	For instance,
	it may be an extra vector-like matter charged under the SM gauge group.
	({\it e.g.}, an extra-quark with a color charge.)
	As a particular case, $\chi$ itself may be chiral SM fields.
	In this case, parametrizing a (flat) direction as $\phi$ which may have a large initial amplitude,
	one finds the same interaction as Eq.~\eqref{yukawa}, given by
	\begin{align}
		{\cal L} = \sum_{i,j}\lambda_{ij} \phi \bar\chi_{{\rm L},i} \chi_{{\rm R},j} +{\rm h.c.},
	\end{align}
	where a summation over gauge indices which are not broken by the
	scalar condensation is promised.
}
First, the $\chi$ fields obtain the $\phi$-dependent mass through this coupling.
It is soon recognized that if $\lambda\phi \gg m_\phi$, where $m_\phi$ is the mass of $\phi$,
the decay process $\phi \to \chi\bar\chi$ is kinematically forbidden except for the region near $\phi\sim 0$.
In this case, the perturbative decay rate at the true vacuum does not apply
and we must take into account particle production phenomena occurring around $\phi\sim 0$
and subsequent thermalization processes of $\chi$.
Moreover, in general, there always be background radiation, or thermal bath, even before the reheating completes.
Then, if $\chi$ is light enough, it is thermalized and obtains a thermal mass.
If the thermal mass is large enough, the $\phi$ decay into $\chi$ is again kinematically forbidden
even if the $\phi$ amplitude is very small.
However, even if the decay process is kinematically forbidden, $\phi$ can dissipate its energy
through $\phi$-$\chi$ scattering processes.
Thermal bath also modifies the effective potential of $\phi$ through thermal effects and hence the dynamics of $\phi$
may be significantly affected.

Therefore, even only the introduction of a simple interaction (\ref{yukawa}) leads to many complicated issues
on the dynamics of $\phi$.
Without careful treatment of these issues, we cannot discuss cosmological effects of $\phi$.
To the best of our knowledge, however, there have not been comprehensive and complete analyses on this topic.
It was only partly attacked by separate literature as follows.

\begin{itemize}

\item
Effective potential of the scalar field in thermal environment is well known
in the case where $\chi$ takes part in thermal bath~\cite{KapustaBook}.
On the other hand, thermal modification on the effective potential when $\chi$ decouples from thermal bath
at large $\phi$ value seems to be less known except for the context of Affleck-Dine mechanism~\cite{Anisimov:2000wx}.
However, this modification is rather a generic feature if $\chi$ has a gauge/Yukawa interaction and needs to be included in the analyses.

\item
Besides thermal modification on the effective potential, $\phi$ also receives thermal dissipation.
Dissipation of oscillating scalar field in thermal bath was extensively studied in the context of warm inflation
in Refs.~\cite{Berera:1995ie,Berera:2008ar,BasteroGil:2009ec},
and also in Refs.~\cite{Yokoyama:2004pf,Drewes:2010pf} in a context of inflation.
Ref.~\cite{BasteroGil:2010pb} also considered the dissipation of scalar fields in a setup close to ours.
However, the dynamical aspects of oscillating scalar field with a large amplitude
were not considered.

\item
The effect of non-perturbative particle production, or called {\it preheating}, was studied in detail in 
Refs.~\cite{Kofman:1994rk,Shtanov:1994ce,Kofman:1997yn} and thereafter.
Most of these studies focused on the $\chi$ particle production neglecting the $\chi$ interaction with thermal bath.
Ref.~\cite{Felder:1998vq} considered the case where produced $\chi$ particles decay into radiation: so-called {\it instant preheating}.
This may efficiently transfers the $\phi$ energy density to radiation even in the case where 
$\phi$ has a large amplitude and the standard calculation of perturbative decay into $\chi$ is not applicable.

\item
In association with thermal modification on the effective potential, the scalar field may fragment into the
non-topological solitons. Once most energy of the oscillating scalar field is absorbed by solitons,
estimation of the dissipation rate and the decay temperature of $\phi$ may be significantly modified.
This is studied in the context of Q-balls when the complex scalar oscillates with an elliptical orbit in the complex plane.
The formation of solitons, however, can occur even in the case of a real scalar.
Its phenomenological consequences have not been investigated in detail in a setup we are interested in.

\end{itemize}

In this paper, we consider the dynamics of $\phi$ having interaction of the type (\ref{yukawa})
in thermal environment.
We take into account various effects listed above in reliable manners
and study in which situation these effects become substantial for the scalar dynamics.
As for the issue of non-topological solitons, we avoid to give definite conclusions
but mention possible consequences.
We only have four model parameters: $m_\phi, \phi_i, \lambda, \alpha (\equiv g^2/4\pi)$,
where $m_\phi$ is the bare mass of $\phi$, $\phi_i$ is the initial amplitude of $\phi$, $\lambda$ is the Yukawa coupling
between $\phi$ and $\chi$,
$g$ is the coupling constant between $\chi$ and thermal bath, which is typically an SM gauge coupling.
In addition, we introduce the reheating temperature $T_{\rm R}$, which controls the density of thermal bath.\footnote{
	In this paper we assume the existence of thermal bath, which comes from the inflaton decay.
	Thus our analysis of the scalar dynamics is not applied to the inflaton itself.
	The inflaton dynamics is also an interesting subject, which will be discussed elsewhere.
	\label{foot:inflaton}
}
Even in this simple setup, there appear many time scales and we must be cautious about
the usage of a particular formalism case-by-case.
Eventually, we have successfully figured out the global picture of the scalar dynamics in broad parameter spaces.
We believe that, although this is a simplified toy model, it captures essential features of the most
applications to realistic models. 
This is because oscillating scalar fields must decay into radiation unless it becomes (a part of) dark matter,
and then it inevitably couples to lighter species, which in turn have interactions with SM particles,
and this is nothing but a situation we are considering.
Even if $\phi$ couples to many fields with coupling strength varying by orders of magnitude,
the dynamics mainly depends on the largest coupling and hence our analyses apply.

This paper is organized as follows.
In Sec.~\ref{sec:over} we describe our basic setup and list some particle physics motivated examples.
In Sec.~\ref{sec:thermal} we review thermal effects on the scalar field evolution
based on the
%the formalism of non-equilibrium quantum field theory. 
closed time path formalism.
In particular, thermal modification on the effective potential and the dissipation rate will be introduced,
which significantly affects the scalar dynamics.
In Sec.~\ref{sec:non}, the effects of non-perturbative particle production are discussed.
This also serves as a dissipation of the oscillation of the scalar field.
In Sec.~\ref{sec:dyn}, we solve the scalar dynamics taking these effects into account.
Readers who are only interested in results may skip preceding sections and only read Sec.~\ref{sec:dyn}.
We conclude in Sec.~\ref{sec:conc}.

%%%%%%%%%%%%%%%%%%%%%%%%%%%%%
\section{Overview} 
\label{sec:over}
%%%%%%%%%%%%%%%%%%%%%%%%%%%%%

%%%%%%%%%%%%%%%%%%%%%%%%%%%%%
\subsection{Setup} 
%%%%%%%%%%%%%%%%%%%%%%%%%%%%%

Let us start with a model where a real scalar $\phi$ couples to another field $\chi$ (fermion) and $\tilde \chi$ (boson) 
at the renormalizable level. The Lagrangian is given by
\begin{equation}
	\mathcal L = \mathcal L_{\rm kin}
	-\frac{1}{2}m_\phi^2\phi^2 - \Lambda\phi \tilde\chi \tilde\chi - \frac{1}{2}h^2\phi^2\tilde\chi^2 
	- \lambda\phi \bar\chi\chi,
	\label{Lb}
\end{equation}
where $\Lambda, h$ and $\lambda$ are coupling constants and $\mathcal L_{\rm kin}$ denotes canonical kinetic terms.
The bare mass for $\chi (\tilde \chi)$ is neglected in what follows.
An additional assumption is that the $\chi (\tilde \chi)$ has gauge or Yukawa couplings,
which is collectively denoted by $g$, to particles in thermal bath.
The introduction of $\chi (\tilde \chi)$ field in (\ref{Lb}) should be regarded as schematic.
Our arguments are not affected even if there are many fields which couples to $\phi$ and 
the interaction term is given by $\mathcal L = \sum_{ij}\lambda_{ij}\phi \bar\chi_i \chi_j$, etc.
For example, if $\chi (\tilde \chi)$ has gauge interactions, summation over the gauge indices should be promised.
Notice that in a non-SUSY theory, there is a large correction to the scalar potential.
We have the Coleman-Weinberg potential~\cite{Coleman:1973jx} as
\begin{equation}
	V_{\rm CW} = \sum_F \epsilon_F \frac{m_\chi^4(\phi)}{64\pi^2}\left[ \ln\frac{m_\chi^2(\phi)}{\mu^2}-\frac{3}{2}\right],
\end{equation}
where $\epsilon_F =+1$ for a real scalar boson $\tilde\chi$, and $\epsilon_F =-2$ for a (Weyl) fermion $\chi$.
This yields the quartic potential of the form $V_{\rm CW} \sim h^4\phi^4\ln\phi, \lambda^4\phi^4\ln\phi$.
It must be included in the analysis of scalar dynamics unless the couplings $h$ and $\lambda$ are small enough to be neglected.

In SUSY, $\phi$ and $\chi$ are components of chiral supermultiplets.
In this case, assuming the superpotential of the form
\begin{equation}
	W = \lambda \phi \chi\bar\chi,
\end{equation}
where we used same symbols for the chiral superfields, we obtain the Lagrangian
\begin{equation}
	\mathcal L = \mathcal L_{\rm kin} - \left(\lambda \phi \chi\bar\chi + {\rm h.c.}\right) 
	-  \lambda^2|\phi|^2\left(|\tilde\chi|^2+|\tilde{\bar\chi}|^2\right)
	- \lambda^2 |\tilde {\chi}|^2 |\tilde {\bar\chi}|^2
	-V_{\rm SB},
\end{equation}
where $\mathcal L_{\rm kin}$ denotes the kinetic terms of the fields,
$\chi (\bar\chi)$ and $\tilde \chi (\tilde{\bar\chi})$ are a fermionic and scalar components of each chiral superfield.
In SUSY, therefore, the Yukawa coupling and the four-point coupling are both given by $\lambda$.
In all the analyses, we assume this for simplicity.
Note that scalars are complex fields.
SUSY breaking effects induce scalar potential for $\phi$ as
\begin{equation}
	V_{\rm SB} = m_\phi^2 |\phi|^2+m_\chi^2 |\tilde\chi|^2+m_{\bar\chi}^2 |\tilde{\bar\chi}|^2
		+ \left( A_\chi \phi \tilde\chi \tilde{\bar\chi} + {\rm h.c.}\right),
\end{equation}
where $m_\phi, m_\chi, m_{\bar\chi}, A_\chi$ are SUSY breaking parameters of same orders of magnitude.
One of the good points for considering SUSY is that radiative corrections to the scalar potential get suppressed
and we do not need to worry much about the Coleman-Weinberg correction, since quartic terms in $\phi$
cancel out and there only remain small logarithmic correction to the quadratic terms from the SUSY breaking effect.
We implicitly assume this in the following.

In general, the motion of scalar condensate in the complex plane becomes elliptical.
However, if the $A$-term contribution to the scalar potential is small, one can approximate the dynamics of
complex scalar field with its one-dimensional radial component.
In the following, we concentrate on this case, and hence the scalar field is reduced to one dimensional
real scalar field.

After all, in order to avoid unnecessary complexity, we consider a following phenomenological model
\begin{equation}
	\mathcal L = \mathcal L_{\rm kin}
	-\frac{1}{2}m_\phi^2\phi^2 - \lambda\phi \bar\chi\chi - \lambda^2 \phi^2\tilde\chi^2 ,
	\label{model}
\end{equation}
where $\phi$ is a real scalar, $\chi$ is a fermion and $\tilde\chi$ is a scalar boson.
We do not include scalar tri-linear interactions because its typical size is much smaller than others.
We study the dynamics of scalar field $\phi$ in the presence of thermal background interacting with $\chi$ and $\tilde \chi$.
We will see that the resulting scalar dynamics in this simple theory is highly non-trivial.
Hereafter, the coupling constant $\lambda$ and the $\chi$ coupling with thermal bath, $g(\equiv\sqrt{4\pi\alpha})$, 
is assumed to be smaller than unity.\footnote{
	For notational simplicity, we often call ``$\chi$'' both for fermion $\chi$ and boson $\tilde\chi$
	unless we need to distinguish them.
}

Let us suppose that $\phi$ has an initial value of $\phi_i$ during inflation
and follow the dynamics of $\phi$.
Conventional arguments are as follows.
After inflation, $\phi$ is frozen at $\phi_i$ until the Hubble parameter decreases to $m_\phi$,
at which $\phi$ begins to oscillate with an initial amplitude of $\phi_i$.
The abundance of $\phi$, in terms of the $\phi$ energy density ($\rho_\phi$) to entropy density ($s$) ratio, 
is then given by
\begin{equation}
	\frac{\rho_\phi}{s} = 
	\begin{cases}
		\displaystyle \frac{1}{8}T_{\rm R}\left( \frac{\phi_i}{M_P} \right)^2 & {\rm if~~}T_{\rm R}<T_{\rm os},\\
		\displaystyle \frac{1}{8}T_{\rm os}\left( \frac{\phi_i}{M_P} \right)^2 & {\rm if~~}T_{\rm R}>T_{\rm os},
	\end{cases}
\end{equation}
where $M_P$ is the reduced Planck scale,
$T_{\rm R}$ is the reheating temperature after inflation and $T_{\rm os} \equiv (\pi^2 g_*/90)^{-1/4}\sqrt{m_\phi M_P}$.
Then, $\phi$ decays into $\chi$ with the decay rate given by
\begin{equation}
\begin{split}
	\Gamma_\phi(\phi\to\bar\chi\chi) &= \frac{\lambda^2 m_\phi}{8\pi} \left(1- \frac{4m_\chi^2}{m_\phi^2} \right)^{3/2}
	.  \label{pert_dec}
\end{split}
\end{equation}
Since $\chi$ is assumed to have sizable couplings to thermal bath, the energy stored in the form of
$\phi$ coherent oscillation goes to thermal bath after the $\phi$ decay.

The above arguments, however, miss some possibly important effects.
First, $\phi$ couples to thermal bath through $\chi$ particles.
Although $\chi$ itself may be heavy enough to decouple from thermal bath at large $\phi$ value,
it may still modify the effective potential of $\phi$.
Second, if the finite-temperature correction to the scalar potential is dominant,
the $\phi$ fluctuation develops to form non-topological solitons.
Once $\phi$ fragments into solitons, their cosmological evolutions are significantly different from 
the case of homogeneous oscillation.
Third, the perturbative decay rate (\ref{pert_dec}) is not simply applied
since $\chi$ can be heavier than $\phi$ during the course of oscillations with large amplitude.
Instead, non-perturbative particle productions when $\phi$ passes through the origin
may be efficient, which in turn reduces the energy of $\phi$ coherent oscillation.

Therefore, even in the simple model like (\ref{model}),
the dynamics of scalar field $\phi$ is complicated.
There appear many time scales related to the above mentioned effects, depending on
the parameters $m_\phi$, $\phi_i$, $\lambda$ and the background temperature $T$.
Without taking into account these effects, one cannot evaluate the 
the abundance of $\phi$ and its cosmological consequences.
In the following sections, we discuss the dynamics of $\phi$ in detail.

It should also be noticed that the following analyses are applied to the more general forms of the
scalar potential for $\phi$.
For example, let us assume that the scalar potential takes the form
\begin{equation}
	V(\phi) = \frac{1}{2}m_\phi^2 \phi^2 -\frac{c}{2}H^2\phi^2 + \frac{\kappa\phi^{n}}{nM_P^{n-4}},
\end{equation}
where $n\geq 4$ is an integer, $H$ is the Hubble parameter and $c$ is an $\mathcal O(1)$ positive constant.
This is the typical potential of the flat direction in the MSSM, if the $A$-term contribution is small enough~\cite{Dine:1995kz}.
In this case, the field sits at $\phi=(cH^2M_P^{n-4}/\kappa)^{1/(n-2)}$ for $H \gtrsim m_\phi$,
and then it begins to oscillate around the quadratic potential with initial amplitude of
$\phi_i=(cm_\phi^2M_P^{n-4}/\kappa)^{1/(n-2)}$ at $H\simeq m_\phi$ if there are no thermal corrections.
It is not difficult to apply our results to such a case.

%%%%%%%%%%%%%%%%%%%%%%%%%%%%%
\subsection{Examples} 
%%%%%%%%%%%%%%%%%%%%%%%%%%%%%

Before going to discuss details of the scalar dynamics, we point out that
broad class of models discussed in particle physics and cosmological contexts 
are actually described by (\ref{model}).

%%%%%%%%%%%%%%%%%%%%%%%%%%%%%
\subsubsection{Curvaton} 
%%%%%%%%%%%%%%%%%%%%%%%%%%%%%

Curvaton is a hypothetical scalar field, which is responsible for the generation of observed
cosmological density fluctuations~\cite{Mollerach:1989hu,Linde:1996gt,Lyth:2001nq,Moroi:2001ct}.
In the simplest case, the curvaton $(\phi)$ has a quadratic potential and is supposed to be
frozen at the initial value $\phi=\phi_i$ during inflation.
Since the curvaton mass $m_\phi$ is much smaller than the Hubble scale during inflation, $H_{\rm inf}$,
it obtains quantum fluctuations characterized by $H_{\rm inf}$,
which in turn seeds the density perturbation with nearly scale-invariant power spectrum.
Since the curvaton must decay in order to convert the fluctuation in the curvaton
into the radiation, we need to introduce couplings between curvaton and some particles, $\chi$, 
which interacts with standard model particles.
Thus the model of (\ref{model}), with/without $\phi^2\tilde\chi^2$ term, describes the `minimal' curvaton model.

An interesting feature of the curvaton model is that it can generate a large (local-type) non-Gaussianity
in the cosmological density perturbation, while it is difficult in the standard single-field inflation model
\cite{Lyth:2005fi,Ichikawa:2008iq,Suyama:2010uj}.
The curvature perturbation $\zeta$ in the curvaton model is given by~\cite{Lyth:2005fi}
\begin{equation}
	\zeta = \zeta_{\rm g}+\frac{3}{5}f_{\rm NL}\zeta_{\rm g}^2.
\end{equation}
Here $\zeta_{\rm g}$ denotes the Gaussian part of $\zeta$ and is given by
$\zeta_{\rm g}=RH_{\rm inf}/(3\pi \phi_i)$ where
\begin{equation}
	R = \left. \frac{3\rho_\phi}{3\rho_\phi+4\rho_r}\right|_{\phi~{\rm decay}},
\end{equation}
which roughly describes the energy fraction of $\phi$ in the Universe at the decay.
The non-linearity parameter, $f_{\rm NL}$, is evaluated by
\begin{equation}
	f_{\rm NL}=\frac{5}{4R}\left(1- \frac{4R}{3}-\frac{2R^2}{3} \right).
\end{equation}
It is clear from this expression that we need $R\lesssim 0.1$ for obtaining
$f_{\rm NL} \sim \mathcal O(10)$.
This means that the curvaton must not dominate the Universe before it decays
in order to obtain large non-Gaussianity.
Therefore, the initial condition should satisfy $\phi_i \ll M_P$, since otherwise
the curvaton eventually dominates the Universe unless the inflaton reheating is so late.
However, as will be seen later, thermal effects are always important if $\phi_i \ll M_P$.
Then the estimates of the curvaton abundance and hence the resulting non-Gaussianity 
are significantly affected.\footnote{
	The estimate of the non-Gaussianity is affected due not only to the change in the curvaton abundance
	but also to the departure of the scalar potential from the quadratic one~\cite{Enqvist:2005pg,Kawasaki:2008mc}.
}
Thus it is important for the curvaton scenario to reconsider the scalar dynamics.

%%%%%%%%%%%%%%%%%%%%%%%%%%%%%
\subsubsection{Flat directions in MSSM} 
%%%%%%%%%%%%%%%%%%%%%%%%%%%%%

Next, let us consider a supersymmetric (SUSY) theory.
In SUSY, there appear many scalar fields as superpartners of the standard model fermions.
It is known that, in the frame work of the minimal SUSY standard model (MSSM),
some combinations of sfermions have vanishing $F$- and $D$-term potential at the renormalizable level
in the SUSY limit~\cite{Affleck:1984fy,Dine:1995kz,Gherghetta:1995dv}.
Such a flat direction in the MSSM, parametrized by a complex scalar $\Phi$, 
obtains scalar potential from the SUSY breaking effect as $V(\Phi) = m_\Phi^2|\Phi|^2$
and also from possible non-renormalizable terms in the superpotential or K\"ahler potential.
The dynamics of flat directions was investigated in detail in the context of Affleck-Dine baryogenesis
\cite{Affleck:1984fy,Dine:1995kz} including thermal effects~\cite{Allahverdi:2000zd,Anisimov:2000wx,Fujii:2001zr}
and the formation of Q-balls~\cite{Coleman:1985ki,Cohen:1986ct,Kusenko:1997si,Enqvist:1997si,Laine:1998rg,Banerjee:2000mb,Kasuya:1999wu,Fujii:2001xp}.
While the motion of flat direction in the complex plane must be elliptical in order to generate sizable baryon number,
it is possible that the motion is nearly one-dimensional if the $A$-term contribution to the scalar potential is somehow small, 
as in the gauge-mediated SUSY breaking models.
Then the dynamics is essentially one dimensional described by the motion of the radial component of $\Phi$.
Since flat directions, which are composed of squarks, slepton and Higgs bosons in the MSSM,
have gauge and Yukawa interactions, the interaction terms like (\ref{model}) appear.
In this case, $\chi$'s are MSSM fields and $\lambda$'s are corresponding gauge or Yukawa couplings.
Thus the dynamics of the flat direction also reduces to a simple model of (\ref{model}).

%%%%%%%%%%%%%%%%%%%%%%%%%%%%%
\subsubsection{Right-handed sneutrino} 
%%%%%%%%%%%%%%%%%%%%%%%%%%%%%

Right-handed neutrinos are often introduced because
it can explain the tiny left-handed neutrino masses through the seesaw mechanism,
and also the observed baryon asymmetry of the universe through the leptogenesis scenario
by their non-equilibrium decay.
In SUSY, there are scalar partners of them, right-handed sneutrinos denoted by $\tilde N_i~(i=1,2,3)$.
The superpotential is given by
\begin{equation}
	W = \frac{1}{2}m_{N_i} N_iN_i + y^N_{ij} N_i L_j H_u,
\end{equation}
where $N_i, L_j$ and $H_u$ denote the right-handed neutrino, left-handed lepton and up-type Higgs superfields,
$m_{N_i}$ is the Majorana masses for the right-handed neutrino and $y^N_{ij}$ is the Yukawa coupling among them.
Focusing on the lightest component, $N_1$, and writing the $D$-flat direction of the $\tilde L_1 H_u$ as $\tilde L_1H_u=\tilde\chi^2$,
the scalar potential is given by
\begin{equation}
	V = |m_{N_1} \tilde N_1 + y^N_{11}\tilde\chi^2 |^2 + V_{\rm SB}.
\end{equation}
Here $V_{\rm SB}$ denotes the SUSY breaking effect including terms such as $Bm_{N_1}\tilde N_1\tilde N_1+{\rm h.c.}$
with a constant $B$ of the soft SUSY breaking scale.
Assuming that initially $\tilde N_1$ has large amplitude and $\tilde \chi=0$,\footnote{
	This assumption is valid if $m_{N_1}\ll H_{\rm inf}$.
}
the scalar potential of the radial component of $\tilde N_1$ is simply quadratic and, 
as a result, the dynamics is described by (\ref{model}).
Cosmological implications of the right-handed sneutrino condensation, mostly in the context of
non-thermal leptogenesis, were discussed in the literature
\cite{Murayama:1992ua,Murayama:1993em,Hamaguchi:2001gw,Hamaguchi:2002vc,Moroi:2002vx,McDonald:2003xq,Allahverdi:2004ix}.
Thermal correction on the right-handed sneutrino dynamics was partly discussed in 
Refs.~\cite{Hamaguchi:2002vc,McDonald:2003xq}, but the complete analyses on this system were not performed so far.
It may be crucial for the estimate of the baryon number generated by the right-handed sneutrino decay.

%%%%%%%%%%%%%%%%%%%%%%%%%%%%%
\subsubsection{Peccei-Quinn scalar} 
%%%%%%%%%%%%%%%%%%%%%%%%%%%%%

The most attractive solution to the strong CP problem is the Peccei-Quinn (PQ) mechanism~\cite{Peccei:1977hh,Kim:1986ax}.
In the so-called hadronic axion model~\cite{Kim:1979if}, the complex PQ scalar field, $\Phi$, couples to 
vector-like quarks $Q$ and $\bar Q$ as $\mathcal L = \lambda \Phi \bar QQ$.
If the PQ scalar obtains a vacuum expectation value of order of $10^9$--$10^{12}$\,GeV, the strong CP problem is solved
without conflict with cosmological/astrophysical observations related with the axion phenomenology.

In SUSY, there is a flat direction in the PQ scalar sector, called saxion.
The saxion feels the scalar potential from the SUSY breaking effect.
Thus the saxion potential at large field value is often quadratic, although this is not mandatory.
The dynamics of the saxion and its cosmological implications were discussed in Refs.~\cite{Asaka:1998ns,Chun:2000jr,
Abe:2001cg,Banks:2002sd,Kawasaki:2007mk,Kim:2008yu,Choi:2009qd,Park:2010qd,Choi:2011rs,Jeong:2011xu,
Kawasaki:2010gv,Kawasaki:2011aa,Nakayama:2012zc,Moroi:2012vu}.
If the initial amplitude of the saxion is much larger than the PQ scale, 
the saxion dynamics resembles the model described by (\ref{model}),
where $\chi$ identified with the PQ quarks, $Q$ and $\bar Q$.
(Thermal correction on the saxion potential in this context was pointed out in Ref.~\cite{Kawasaki:2010gv,Moroi:2012vu}.)
One should notice that the final stage of the PQ scalar dynamics toward the symmetry breaking minimum 
may include additional steps.
For example, if the saxion is trapped at the origin due to the particle production effect,
the saxion finally causes thermal inflation~\cite{Yamamoto:1985rd,Lyth:1995hj}.
However, the first stage of the saxion dynamics still falls into the model of (\ref{model}).
Details of the saxion dynamics certainly depend on how the PQ scalar is stabilized, 
and more complete analysis will be presented elsewhere.

%%%%%%%%%%%%%%%%%%%%%%%%%%%%%
\section{Thermal effects on the scalar field dynamics} 
\label{sec:thermal}
%%%%%%%%%%%%%%%%%%%%%%%%%%%%%

In this section we discuss thermal effects on the dynamics of homogeneous 
scalar condensation in this model.
Since the $\phi$ field interacts with thermal plasma through $\chi$ fields, $\phi$'s dynamics may be affected.
In particular, there are two effects caused by thermal plasma:
thermally modified potential for $\phi$ and the dissipation of $\phi$ to thermal plasma.
Thermal effects are qualitatively different whether the field value is large or small:
$\lambda \phi (t) \gg T$ or $\lambda \phi (t) \ll T$.
Hence, we study these regimes separately in the following.
Basic ingredients for closed time path formalism are summarized in Appendix \ref{sec:CTP} briefly.
An expectation value of operator $\hat{A}$ with respect to the physical state $\hat\rho$ is denoted by
\begin{align}
	\< \hat{A} \> := {\rm tr} \{ \hat\rho \hat{A} \}.
\end{align}
In the following, we assume that the system has  a spacial translational invariance
and the homogeneous condensation of $\hat\varphi$ is denoted as $\phi = \< \hat\varphi \>$.

%%%%%%%%%%%%%%%%%%%%%%%%%%%%%
\subsection{Large Field Value Regime}  \label{sec:large}
%%%%%%%%%%%%%%%%%%%%%%%%%%%%%

First, let us study the large field value regime: $\lambda \phi (t) \gg T$.
We separate the relevant time scales and follow a
coarse-grained effective motion equation for $\phi$.

As shown a posteriori, a typical dynamical scale of scalar condensation $\phi$ at this regime is
much slower than the thermalization time scale: $m_{\rm eff}^\phi \ll \Gamma_{\rm th}$, 
if the $\phi$ oscillates with the thermally modified potential.
Of course, this condition is automatically satisfied if the $\phi$ oscillates with the vacuum mass
$m_\phi$ with $m_\phi \ll \Gamma_{\rm th}$.
Then we can assume that the ``fast'' fields in thermal bath feel the ``slow'' $\phi$'s 
dynamics as almost static. %namely quasi stationary. 
Hence, on the typical time scale of $\phi$'s dynamics, 
the other fields in thermal bath have long enough time to thermalize with an each value of 
background field $\phi$ and completely forget about their past.
In such a regime, 
%we can see the thermal plasma as thermal equilibrium with the background field $\varphi$, and then
the equation of motion can be simplified by
tracing out all the degrees of freedom in thermal equilibrium.

There is one more important aspect at this regime. 
Since the field value is large $\lambda \phi (t) \gg T$,
the number densities of $\chi$ particles are very small.
%compared to those of typical light fields in thermal bath. 
Therefore $\phi$ does not feel the existence of these particles directly.
However, this does not immediately mean that %we can treat 
the $\phi$'s dynamics is completely
free from thermal plasma, despite the fact that the interaction between 
$\phi$ and thermal bath is only mediated by the heavy $\chi$ fields.
This is because %$\phi$'s variation does affect 
the free energy of thermal plasma depends on the background $\phi$ field
at a higher loop order, and hence the ``pressure''\footnote{
Analogous to
$
  P = - \fr{\der F}{\der x}
$
in statistical mechanics.
}  dominantly affects the $\phi$'s dynamics
if the vacuum mass $m_\phi$ is very small.

With the above consideration in mind, let us derive the effective equation of motion of
$\phi$'s homogeneous condensation.
Initially, the system is prepared as the canonical ensemble 
with the non-vanishing background field expectation value $\phi(t_i)$.
In principle, one may follow all the dynamics in terms of equations of motion derived
from Closed Time Path 2PI (nPI) effective action:
$
0 = \delta \Gamma [\phi, G_{ij}]/ \delta \phi
$,
$
0 = \delta \Gamma [\phi, G_{ij}]/ \delta G_{ij}
$
where the subscript $i$ runs all the fields, but it is practically difficult
(See Refs.~\cite{Berges:2004yj,CalzettaHuBook} and references therein).
Since all the fields except $\phi$ remain in thermal equilibrium  
with the background field $\phi (t)$, we can safely assume that 
the propagators $G_{ij}$ are given by thermal ones.
Therefore, the effective equation of motion is reduced to
\begin{align}
  0 = \fr{\delta \Gamma [\phi]}{\delta \phi} 
  = \fr{\delta S_\phi}{\delta \phi} + \fr{\delta \tilde{\Gamma}}{\delta \phi}
\end{align}
where $S_\phi$ is the $\phi$'s classical action and
$\tilde{\Gamma}$ denotes the sum of bubbles 
calculated in terms of the thermal propagators
with the background field $\phi$.

Let us evaluate $\tilde{\Gamma}$ approximately.
Since the motion of $\phi$ is very slow compared to the thermal plasma,
we will neglect the time dependence of $\phi$ as a first step,
and then take it into account approximately.
%The dominant contribution is obtained with the dynamics of $\phi$ being 
%completely neglected. 
If the $\phi$ is regarded as static, 
$\tilde{\Gamma}$ is merely the ``free energy'' of thermal plasma with the constant 
background field $\phi$, 
and hence we have $\delta \tilde{\Gamma} / \delta \phi \simeq - \der {\cal V}_{\rm eff}/ \der \phi$.
Since the $\chi$ particles are absent at $T \ll \lambda \phi (t)$ due to the Boltzmann suppression, 
we can integrate out $\chi$ fields first 
in calculation of the free energy. This leads to the effective operator
which contributes to the running gauge coupling constant $g$\footnote{
 	In general, the $\chi$'s large mass from the $\phi$'s field value 
	affects the running Yukawa coupling,
	for instance, if the $\chi$ field interacts with other light degrees of freedom via Yukawa interaction, or
	if the $\chi$ field mixes to the SM fermions, which have Yukawa interaction.
	Throughout this paper, we assume that the gauge coupling contributes dominantly for simplicity.
}
\begin{align}
  \fr{A}{16 \pi^2} \ln (\lambda^2 \phi^2/T^2) \ F^{a\,\mu \nu} F^a_{\mu \nu}
\end{align}
where $A$ is a constant determined by the representation of $\chi$ fields.
Because the free energy of hot QCD plasma has a contribution proportional to 
$g^2(T)\, T^4$,
this term induces the so-called thermal logarithmic potential~\cite{Anisimov:2000wx}:
\begin{align}
  {\cal V}_{\rm eff} \supset a \, \alpha (T)^2 \,  T^4 \ln (\lambda^2 \phi^2/T^2)
\end{align}
where $a$ is an order one constant.\footnote{
	The sign of the coefficient $a$ depends on the model.
	In the model we are considering, $\chi$ is a matter field, which is either a fermion or a scalar boson
	having gauge interactions. In this case, $a$ is positive.
	On the other hand, if $\chi$ is a gauge boson, $a$ can be negative.
	In all the analyses in this paper, we assume that $a$ is positive.
}

The typical dynamical scale of this term is given by 
$m^\phi_{\rm eff}{^2} \sim \alpha^2 T^4 / \phi^2$.
Thus, if the $\phi$ oscillates with the thermal logarithmic potential,
the motion of $\phi$ is adiabatic with respect to the typical thermalization time scale
of thermal plasma, $\Gamma_{\rm th} \sim \alpha T \gg m^\phi_{\rm eff}$,
at the large field value regime, $T \ll \lambda \phi (t)$.

Next, let us take into account the time dependence of $\phi$ at the leading order, and
derive the dissipative coefficient, which describes the typical relaxation 
time scale of $\phi$. We will separate the $\phi (t)$ field
into the constant background $v$ at time $t$ and the small deviation $\delta \phi(t)$,
and then evaluate $\tilde\Gamma$ up to the first order in $\delta\phi (t)$.
$\tilde{\Gamma}$ can be expanded as~\cite{Moss:2006gt}
\begin{align}
  \fr{1}{V}\fr {\delta \tilde{\Gamma} [v + \delta \phi]}{\delta \phi (t)} &= 
  \fr{1}{V}
  \sum_{n} \fr{1}{n\ !} \int dt_1 \cdots dt_n\
  \left. 
  \fr{\delta^{n+1} \tilde{\Gamma}}{ \delta \phi (t) \delta \phi (t_1) \cdots \delta \phi (t_n)} \right|_{\phi=v}
  \delta \phi (t_1) \cdots \delta \phi (t_n) \\
  &= - \fr{\der {\cal V}_{\rm eff}}{\der \phi} - \int d\tau\ \Pi_{\rm ret} (t- \tau,{\bf 0}) \ \delta \phi(\tau)
  + \cdots \label{eq:eom_s}
\end{align}
up to the first order in $\delta \phi$. 
Here $V$ denotes the spacial volume.
The first term is the free energy, which is already obtained above.
At the leading order, the self energy is given by
\begin{align}
  \Pi_{\rm ret} (x) & := - i \theta (x^0) \Pi_J (x); \ \ \
   \Pi_J (x) = \< [ \hat{O} (x), \hat{O} (0) ]  \>
\end{align}
where the effective interaction, obtained from integrating out $\chi$ fields, is given by
\begin{align}
	{\cal L}_{\rm int} = 
	\delta \phi \, \hat{O}; \ \ \ 
	\hat{O} :=
	\fr{A}{8 \pi^2v} F^{a\,\mu \nu} F^a_{\mu \nu}. \label{eq:op_trace}
\end{align}
Here the ensemble average $\< \cdots \>$ is merely the thermal one,
because the thermal plasma remain in thermal equilibrium.
%we denoted the canonical ensemble average as $\< \cdots \>_T$.
Since the relevant time scale of the self energy is determined by thermal degrees of freedom,
it is much faster than that of $\phi$'s dynamics.
Then we can approximate $\delta \phi (\tau)$ as $\delta \phi (\tau) \simeq \dot{\phi}(t) (\tau-t)$,
and the following equation is obtained
\begin{align}
   \int d\tau \ \Pi_{\rm ret} ( t - \tau,{\bf 0})\ \delta \phi (\tau)
   &= -\,  i \int d\tau\ \Im \Pi_{\rm ret} (\tau,{\bf 0})\ \tau \ \dot{\phi}(t) \nn
   %&= - \left. \fr{\der \Im \Pi_{\rm ret} (\omega,{\bf 0})}{\der \omega} \right|_{\omega = 0} \dot{\phi} (t)\\
   &= - \lim_{\omega \rightarrow 0} \fr{\Im \Pi_{\rm ret} (\omega, {\bf 0})}{\omega} \ \dot{\phi}(t).
   \label{eq:diss_coef}
\end{align}
In the first equality, we have used the fact that the real and imaginary part of 
$\Pi_{\rm ret}(t,{\bf 0})$ are even and odd functions in $t$ respectively.
As can be seen from Eq.~\eqref{eq:diss_coef},
the dissipative coefficient is imprinted in the imaginary part of self energy:
\begin{align}
  \Gamma_\phi := - \lim_{\omega \rightarrow 0} \fr{ \Im \Pi_{\rm ret} (\omega, {\bf 0})}{\omega}
  =  \lim_{\omega \rightarrow 0} \fr{ \Pi_{J} (\omega, {\bf 0})}{ 2 \omega}. \label{eq:def_diss_coef}
\end{align}
In the second equality, we have used the Kramers-Kronig relation.
As a result, the dissipation coefficient can be obtained from
\begin{align}
	\Gamma_\phi
	=
	\lim_{\omega \rightarrow 0} \frac{1}{2 \omega} \int d^4 x\, e^{i\omega t}
	\< [\hat O (t,{\bf x}), \hat O (0, {\bf 0})] \>; \ \ 
	\hat O
	= \frac{A}{8 \pi^2 v} F^{a\, \mu\nu} F^a_{\mu \nu}. \label{eq:diss_coef_fin}
\end{align}

Note that the operator $\hat{O}$ is related to the trace anomaly of gauge field.
As pointed out in Refs.~\cite{Bodeker:2006ij,Laine:2010cq}, 
the dissipative coefficient induced by the trace anomaly
is directly related to the bulk viscosity of hot QCD plasma~\cite{Arnold:2006fz}:
\begin{align}
  \zeta = \frac{1}{9} \lim_{\omega \rightarrow 0} \frac{1}{2 \omega}
  \int d^4 x \ e^{i \omega t} \< [ T^\mu{_\mu} (t,{\bf x}),T^\nu{_\nu} (0,{\bf 0})  ] \>,
\end{align} 
where
\begin{align}
  T^\mu{_\mu} (t,{\bf x}) \simeq 
  - \frac{b_0}{2} F^{a\, \mu \nu} F^a_{\mu\nu}, %; \ \ \ b_0 :=  \frac{11 N}{3 (4\pi)^2}. 
\end{align}
with $b_0$ defines the $\beta$ function of gauge coupling.
As can be seen from Eq.~\eqref{eq:diss_coef_fin}, the dissipative coefficient can be expressed as~\cite{Bodeker:2006ij,Laine:2010cq},
\begin{align}
  \Gamma_\phi
  = 
  \( \frac{A}{8 \pi^2} \)^2 
  \frac{36\, \zeta}{b_0^2 v^2}
  \sim
  \( \frac{A}{8 \pi^2} \)^2 \frac{( 12 \pi \alpha )^2}{ \ln \alpha^{-1}}\ \frac{T^3}{v^2}. \label{eq:diss_coef_bulk}
\end{align}
Here the bulk viscosity $\zeta$ is evaluated
at the weak coupling regime~\cite{Arnold:2006fz}.~\footnote{
Here we assumed $m_q \ll \alpha T$
where $m_q$ is the heaviest zero temperature quark mass in thermal bath.
  ``In thermal bath'' means that the zero temperature mass is at most $T$.
} 

Finally, with taking account of the adiabatic expansion of the universe $H \ll \Gamma_{\rm th}$,
we obtain the effective equation of motion of $\phi$ at the large field value regime
from Eqs.~\eqref{eq:eom_s} and \eqref{eq:diss_coef_bulk}:
\begin{align}
  \ddot{\phi} + \( 3 H + \Gamma_\phi \) \dot{\phi} + m_\phi^2 \phi
  + {\cal V}'_{\rm eff} = 0 \label{eq:eom_ad}
\end{align}
where the dissipative coefficient and the effective potential are given by
\begin{align}
  &\Gamma_\phi
  \sim  \
  \( \frac{A}{8 \pi^2} \)^2 \frac{( 12 \pi \alpha(T) )^2}{ \ln \alpha(T)^{-1}}\ \frac{T^3}{\phi^2} \\
  &{\cal V}_{\rm eff} (\phi)
  \simeq  \ a \, \alpha (T)^2 \, T^4 \ln ( \lambda^2\phi^2/T^2)
\end{align}
respectively. Here we have omitted the constant term of effective potential 
which is independent of the $\phi$ at the large field value regime.

%%%%%%%%%%%%%%%%%%%%%%%%%%%%%
\subsection{Small Field Value Regime}
\label{sec:small}
%%%%%%%%%%%%%%%%%%%%%%%%%%%%%

Second, let us study the small field value regime: $\lambda \phi (t) \ll T$.
In this regime, $\chi$'s number density can not be neglected and the $\chi$ particles 
in thermal plasma may directly affect the $\phi$'s dynamics.
In the following, we assume that the dynamics of $\phi$ is not so violent 
as $\chi$'s number density cannot remain 
the Bose-Einstein distribution.
In other words, the $\chi$'s propagators can be well approximated with thermal propagators.
Typically, this is the case where $\phi$'s dynamics is slow enough for thermal 
plasma to remain thermal equilibrium.
A possible non-perturbative production when the $\phi$ passes through the origin
is discussed in the next Sec.\ \ref{sec:non}
and the applicability of discussion given in this section is clarified.

In the small field value regime, the effective equation of motion can be expressed as
\begin{align}
  \ddot{\phi}(t) +3H \dot{\phi}(t) +  m_\phi^2 \phi(t) +
  \int d\tau\ \Pi_{\rm ret} (t-\tau,{\bf 0})\, \phi (\tau) = 0. \label{eq:eom_small}
\end{align}
The self energy can be decomposed into the local and non-local parts as
\begin{align}
&  \Pi_{\rm ret} (t,{\bf 0}) = M^2 \delta (t) + \tilde{\Pi}_{\rm ret} (t,{\bf 0}),
\end{align}
where
\begin{align}
 & M^2 =\ 2 \lambda^2  \< \hat{\tilde\chi}^2 \>, \\
 & \tilde{\Pi}_{\rm ret} (t,{\bf 0}) = - i \theta (t) \tilde{\Pi }_J (t,{\bf 0}); \ \ \ 
  \tilde{\Pi}_J (t,{\bf x}) = \< [\hat{O}(t,{\bf x}), \hat{O}(0,{\bf 0}) ] \>,
\end{align}
at the leading order in $\lambda$ and $\Lambda$,
with the operator $\hat{O}$ being 
\begin{align}
  \hat{O} = \lambda \bar{\chi} \chi + \Lambda \tilde\chi \tilde\chi.
\end{align}
Here the ensemble average $\< \cdots \>$ is also the thermal one.\footnote{
	Although we do not consider scalar tri-linear coupling in the application in the following sections,
	here we give formulae for the case of scalar tri-linear coupling just for illustration.
}

Note that
if there exist several $\chi$s (for instance, these are charged under some gauge group of SM),
the degrees of freedom should be multiplied to all the following results.

The self energy is computed perturbatively in terms of thermal propagators of $\chi$ fields.\footnote{
	See Appendix.~\ref{sec:thermal_eq} for explicit forms of thermal propagators.
} 
First, let us evaluate the local part and the dominant real part of self energy, which lead to
the thermal mass proportional to $T^2$.
The local part from one real bosonic $\tilde\chi$ is given by~\cite{Dolan:1973qd}
\begin{align}
  M^2 = & \ 2 \lambda^2 \int \frac{d^3 k}{ (2\pi)^3} \frac{f_B (\omega_{\bf k})}{\omega_{\bf k}} 
  \simeq \frac{\lambda^2 T^2 }{6},
\end{align}
where $\omega_{\bf k} = \sqrt{\lambda^2 \phi^2 + {\bf k}^2}$,
at the leading order in $\lambda \phi / T$, with $f_B$ being the Bose-Einstein distribution. 
Similarly, the dominant real part of self energy
from two Weyl fermions $\chi_{\rm L}$ and $\chi_{\rm R}$ 
via the Yukawa interaction can be evaluated as
\begin{align}
  \Re \tilde{\Pi}_{\rm ret}  \simeq &\ \frac{\lambda^2 T^2}{6}
\end{align}
at the leading order in $\lambda\phi/T$~\cite{Thoma:1994yw}.
Note that if the field value is large, then these ``thermal masses'' of $\phi$ are absent 
due to the Boltzmann suppression of $\chi$ particles.
Hereafter thermal masses are denoted as $m_{\rm th}^i (T)$ collectively
where the superscript $i$ denotes the species: $\phi, \chi$.

Next, let us evaluate the dissipative coefficient imprinted in the imaginary part of self energy.
The dissipative coefficient is given by
\begin{align}
  \Gamma_\phi =  - \left. \frac{\Im \tilde{\Pi}_{\rm ret} (\omega,{\bf 0}) }{\omega} 
  \right|_{\omega = m^\phi_{\rm eff}} 
  = \left. 
    \frac{ \tilde{\Pi}_J (\omega,{\bf 0}) }{2\omega}
  \right|_{\omega = m^\phi_{\rm eff}},
\end{align}
where the effective mass of $\phi$, $m^\phi_{\rm eff}$, is given by 
\begin{align}
  m^\phi_{\rm eff} = \sqrt{m_\phi^2 + m_{\rm th}^\phi(T)^2}
\end{align}
at this regime.
If the effective mass is small compared to the typical thermalization time scale,
the dissipative coefficient can be well approximated with $m^\phi_{\rm eff} \rightarrow 0$,
as in the case of the oscillation with the thermal logarithmic potential.
The self energy can be expressed as
\begin{align}
  \tilde{\Pi}_J (m_{\rm eff}^\phi,{\bf 0})
  = & \
  \lambda^2 \int \frac{d^4 q}{(2\pi)^4} \( f_F (q_0) - f_F (q_0 + m_{\rm eff}^\phi) \)
  {\rm tr }\left[ \rho_\chi^F (q_0,{\bf q}) \rho_\chi^F (m^\phi_{\rm eff}+q_0,{\bf q}) \right] 
  \label{eq:self_fermion}\\
  &+
  \Lambda^2 \int \frac{d^4 q}{\( 2\pi \)^4} \( f_B(q_0) - f_B (q_0 - m_{\rm eff}^\phi) \)
  \rho_\chi^B (q_0,{\bf q}) \rho_\chi^B (m^\phi_{\rm eff}-q_0,{\bf q}), \label{eq:self_boson} 
\end{align}
where $\rho_\chi^{B/F}$ is the spectral density for the bosonic/fermionic $\chi$ field
and $f_{B/F}$ is the Bose-Einstein/Fermi-Dirac 
distribution respectively.

For simplicity, we will assume that the spectral density is well approximated by
the Breit-Wigner form. The Breit-Wigner form of spectral density for boson is given by
\begin{align}
  \rho^B_\chi (q_0,{\bf q}) = & \
  \frac{2 q_0 \Gamma_{\bf q}}
  {[ q_0^2 - \Omega_{\bf q}^2 ]^2 + [q_0 \Gamma_{\bf q}]^2 }
\end{align}
where $\Omega_{\bf q} = \sqrt{m_{\rm th}^{\chi,B}(T)^2 + \lambda^2 \phi^2 + {\bf q}^2 }$.
And the Breit-Wigner form for fermion is given by~\cite{Wang:1999mb}
\begin{align}
  \rho^F_\chi (q_0,{\bf q}) = & \
  \sum_{s = \pm}
  \frac{Z_{\bf q}^s}{2}
  \left[ 
  \frac{\Gamma_{\bf q}^s}{[q_0 - \Omega_{\bf q}^s]^2 + \Gamma_{\bf q}^s{^2}/4}
  \( \gamma_0 - \hat{\bf q} \cdot  \mbox{\boldmath{$\gamma$}} \)
  +
  \frac{\Gamma_{\bf q}^s}{[q_0 + \Omega_{\bf q}^s]^2 + \Gamma_{\bf q}^s{^2}/4}
  \( \gamma_0 + \hat{\bf q} \cdot  \mbox{\boldmath{$\gamma$}} \)
  \right]. \label{eq:spec_fermion}
\end{align}
Here the plus $(s=+)$ and minus $(s=-)$ contribution correspond 
to the ordinary particle like excitation in vacuum and
the new collective excitation in thermal plasma, so called the plasmino, respectively~\cite{Weldon:1982bn,Kiessig:2011fw}.
$\Omega_{\bf p}$ is the quasi-particle energy and $\Gamma_{\bf p}$ corresponds to 
the typical relaxation time scale of quasi-particle, so called the thermal width.
In the following, for simplicity, we will neglect the plasmino contribution and
the dispersion relation is approximated by 
$\Omega^+_{\bf p} \simeq \sqrt{m_\infty^{\chi,F} (T)^2 + {\bf p}^2} $
where $m_\infty^{\chi,F}$ is the asymptotic mass, which is given by 
$m^{\chi,F}_\infty = \sqrt{2} m^{\chi,F}_{\rm th}$~\cite{Kiessig:2011fw}.
Note that Eq.~\eqref{eq:spec_fermion} is valid at the very small $\phi$'s field value regime.
If the $\phi$'s field value is not so small $T \gg \lambda \phi \gg gT$,
then the spectral density of fermion can be approximated with~\cite{BasteroGil:2010pb}
\begin{align}
  \rho^F_\chi (q_0,{\bf q})
  =
  \( \lambda \phi + \Slash{q} \) \
  \frac{2 q_0 \Gamma_{\bf q}}{ [q_0^2 - \omega_{\bf q}^2 ]^2 + [q_0 \Gamma_{\bf q}]^2 }, 
  \label{eq:spec_fer_dirac}
\end{align}
where $\omega_{\bf q} = \sqrt{\lambda^2 \phi^2 + {\bf q}^2}$.

To evaluate Eqs.\ \eqref{eq:self_boson} and \eqref{eq:self_fermion},
let us consider typical two cases: 
(i) the $\phi$'s dynamics can be regarded as adiabatic with respect to thermal plasma, and
(ii) the $\phi$'s amplitude is small $\lambda \tilde\phi \ll gT$~\footnote{
	In this case, the efficient non-perturbative particle production is absent.
	See also Sec.\ \ref{sec:non}
}
(not necessarily $m_{\rm eff}^\phi \ll \alpha T$).

In the first case (i), $m_{\rm eff}^\phi \ll \alpha T$, 
the dissipative coefficient can be approximated with~\cite{BasteroGil:2010pb}
\begin{align}
  \Gamma_\phi
  \simeq & \
  \frac{\lambda^2}{2 T} \int \frac{d^4 q}{(2\pi)^4} f_F (q_0)\( 1 - f_F (q_0) \)
  {\rm tr }\left[ \rho_\chi^F (q_0,{\bf q}) \rho_\chi^F (q_0,{\bf q}) \right] \label{eq:diss_fermion} \\
  & + \
  \frac{\Lambda^2}{2 T} \int \frac{d^4 q}{\( 2\pi \)^4} f_B(q_0)\( 1 + f_B (q_0)\)
  \rho_\chi^B (q_0,{\bf q}) \rho_\chi^B (q_0,{\bf q}) \label{eq:diss_boson} 
\end{align}
Using the Breit-Wigner form for the spectral density, one finds the dissipative coefficient 
from Eq.~\eqref{eq:diss_boson} as
\begin{align}
  \Gamma^B_\phi \sim 
  \frac{\Lambda^2}{ \alpha T },
\end{align}
where the thermal width is roughly approximated by $\Gamma_{\bf q} \sim \alpha T$.
On the other hand,
the dissipative coefficient from Eq.\ \eqref{eq:diss_fermion} can be roughly evaluated as
\begin{align}
  \Gamma^F_\phi \sim
  \begin{cases}
     \lambda^2 \alpha T & {\rm for~~}  \ \lambda \phi \lesssim  \alpha T \\
     \lambda^2 \cfrac{\lambda^2 \phi^2}{\alpha T} & {\rm for~~} \ \alpha T \lesssim  \lambda \phi \ll T,
  \end{cases}
\end{align}
where the thermal width is roughly approximated by $\Gamma_{\bf q} \sim \alpha T$.
Note that $\lambda^2 \tilde\chi^2 \phi^2$ term also leads to the same order contribution as
the latter one.
In addition, 
note that we simply extrapolate the obtained dissipation coefficient to 
the intermediate regime, and hence
the result in the interval between two regimes, $\lambda\phi\sim \alpha T$, 
is a rough approximation,
and that the latter expression is not applicable at $\lambda \phi \sim T$ since
the exponential suppression factor in the integrand dominates.
As pointed out in Ref.~\cite{Berera:1998gx}, above {\it one-loop} results are merely approximate 
ones because higher-loop contributions such as ladder diagrams are comparable to {\it one-loop}
ones with the vanishing external energy $\omega \rightarrow 0$. 
This is much like what happens in the calculation of viscosity coefficients 
from Kubo formulas~\cite{Jeon:1994if,Arnold:2006fz}. As discussed in~Ref.~\cite{BasteroGil:2010pb},
the resummation of infinitely many diagrams can change the {\it one-loop} result by several factors.
In the following, however, we will estimate the dissipative coefficient with {\it one-loop} diagrams
as a rough approximation.

In this case (i), the obtained equation is formally equivalent to Eq.\ \eqref{eq:eom_ad},
since the free energy has the thermal mass term $\lambda^2 T^2 \phi^2 $
at the small field value regime.

In the second case (ii), the dissipative coefficient is easily obtained
if the ``decay'' of $\phi$ to $\chi$ is kinematically allowed~\cite{Drewes:2010pf}:
\begin{align}
  \Gamma^B_\phi \simeq &\
  \frac{\Lambda^2}{16 \pi m_{\rm eff}^\phi} 
  \sqrt{ 1- 4\frac{m_{\rm th}^{\chi,B} (T)^2 }{m^\phi_{\rm eff}{^2} } }
  \( 1 + 2 f_B (m^\phi_{\rm eff}/2) \)
  \theta ( m_{\rm eff}^\phi - 2 m_{\rm th}^{\chi, B} (T) ) \label{eq:small_dis_b} \\
  \Gamma^F_\phi \simeq & \
  \frac{\lambda^2 m^\phi_{\rm eff}}{8 \pi} \sqrt{1 - 4 \frac{m_\infty^{\chi,F}(T)^2}{m^\phi_{\rm eff}{^2}} } 
  \( 1 - 2 f_F (m^\phi_{\rm eff}/2) \)
  \theta ( m_{\rm eff}^\phi - 2 m_\infty^{\chi, F}(T) ). \label{eq:small_dis_f}
\end{align}
If the decay is kinematically forbidden $ ( \alpha T \ll )\, m_{\rm eff}^\phi \ll gT$, 
then the dissipative coefficient is from the tail of Breit-Wigner distribution due to the thermal width, 
and hence it is suppressed by the coupling in thermal plasma $g$, compared to the 
``decay''~\cite{Yokoyama:2005dv,Drewes:2010pf}.

%%%%%%%%%%%%%%%%%%%%%%%%%%%%%
\section{Non-perturbative particle production} 
\label{sec:non}
%%%%%%%%%%%%%%%%%%%%%%%%%%%%%

Let us discuss the effects of non-perturbative particle production in this model.
Since the scalar field $\phi$ passes through the origin during the coherent oscillation,
the adiabaticity of the coupled particles $\chi$ is necessarily broken.
It implies that the $\chi$ particles are produced in each oscillation 
even if the perturbative decay of $\phi$ is not efficient.
This is called the preheating~\cite{Kofman:1994rk,Kofman:1997yn}.
For a boson $\tilde\chi$, the production efficiency is enhanced
as the $\tilde\chi$ particle number increases and this leads due to the parametric resonance effect
if the dissipative effect of $\chi$ is not large~\cite{Kasuya:1996aq}.
Even for a fermionic $\chi$, the preheating can have a significant effect
on the reduction of the energy density of $\phi$ through the instant preheating~\cite{Felder:1998vq},
if $\chi$ has sizable interaction with other particles.
(See Refs.~\cite{Greene:1998nh,Giudice:1999fb} for a theory of fermionic preheating.)

%%%%%%%%%%%%%%%%%%%%%%%%%%%%%
\subsection{The case of zero temperature mass} 
\label{sec:non-zero}
%%%%%%%%%%%%%%%%%%%%%%%%%%%%%

First, we consider the case where the $\phi$ oscillates with the zero-temperature mass
and its time dependence is expressed as $\phi(t) = \tilde \phi \sin (m_\phi t)$ in one oscillation.
%This is the case if $m_\phi > \alpha^2 T_{\rm R}^2M_P/\phi_i^2$.
%Then $\phi$ oscillates with a frequency of $m_\phi$ around the origin.
Most discussion below applies for both fermion $\chi$ and boson $\tilde\chi$ as long as their coupling constants to $\phi$ are same
(see (\ref{model})), and we do not distinguish them unless otherwise stated.
The mass of $\chi$ varies with time and it has a frequency\footnote{
	If $\chi$ obtains a large mass from other sources, such as large VEV of flat direction in the MSSM,
	the efficiency of preheating is significantly reduced~\cite{Allahverdi:2005mz,Allahverdi:2007zz}.
	We do not consider such a case.
}
\begin{equation}
	\omega_\chi^2 = k^2 + m^\chi_{\rm th}(T)^2 + \lambda^2 \phi(t)^2,
\end{equation}
where $k$ is a wavenumber and $m^\chi_{\rm th}(T)$ denotes the thermal mass of $\chi$, given by
\begin{equation}
	m^\chi_{\rm th} (T) \sim gT, \label{mchiT}
\end{equation}
Notice that it is not evident that we can take $m^\chi_{\rm th} \sim gT$
since the $\phi$ oscillation frequency may be larger than the $\chi$ thermalization rate.
However, we will see that in the practical use, Eq.~(\ref{mchiT}) gives appropriate results.

Particle production occurs if the following adiabaticity condition, $|\dot\omega_\chi/\omega_\chi^2|\ll1$,
is violated~\cite{Kofman:1997yn}, where we have
\begin{equation}
	\left|\frac{\dot\omega_\chi}{\omega_\chi^2}\right| = \frac{\lambda^2 \tilde\phi^2 m_\phi \sin(m_\phi t)\cos(m_\phi t)}
	{[k^2+m^\chi_{\rm th}(T)^2+\lambda^2\tilde\phi^2\sin^2(m_\phi t)]^{3/2}}.
\end{equation}
First, consider the case of $m^\chi_{\rm th}(T) \ll m_\phi$.
The typical wave number $k_*$, below which the $\chi$ mode is amplified, is given by $k_*=\sqrt{\lambda m_\phi \tilde\phi}$.
The typical time interval $\Delta t_*$, in which the adiabaticity is temporary violated, is estimated to be $\Delta t_*\sim 1/k_*$.
This is much smaller than the oscillation period for $m_\phi \ll \lambda \tilde\phi$. 
%We always assume this in the following discussion.
The created $\chi$ particle number density during the passage of the minimum $\phi\sim 0$ is given by
\begin{equation}
	n_\chi \simeq \frac{k_*^3}{8\pi^3} = \frac{(\lambda m_\phi \tilde\phi)^{3/2}}{8\pi^3}.   \label{nchi}
\end{equation}
Next, consider the opposite case: $m^\chi_{\rm th} (T) \gg m_\phi$.
The condition $|\dot\omega_\chi/\omega_\chi^2|\gg1$ requires $k_*^2 \gg m^\chi_{\rm th}(T)^2$, 
which is rewritten as
$\lambda\tilde\phi \gg m^\chi_{\rm th} (T)^2/m_\phi$.
As long as this condition is satisfied, a typical time interval, in which the particle production occurs, 
is much smaller than the oscillation period: $\Delta t_* \ll 1/m_\phi$.
Thus we impose the following condition for efficient particle production:
\begin{equation}
	\lambda \tilde \phi \gg {\rm max} \left\{ m_\phi, \frac{g^2T^2}{m_\phi} \right\}.  \label{phimax}
\end{equation}
If this condition is met, modes with $k \lesssim k_*$ is amplified and the produced number density is
estimated by Eq.~(\ref{nchi}).
Note that this implies that if the amplitude is very small $\lambda \tilde\phi \ll gT$,
the efficient non-perturbative particle production does not occur.
%Note that this roughly implies $\lambda \tilde\phi \gg gT$ and hence efficient particle production does not occur
%in the regime where $\chi$ is always thermalized during the $\phi$ oscillation.

Subsequent evolution of the system crucially depends on the decay/dissipation rate of $\chi$: $\Gamma_\chi$~\cite{Kasuya:1996aq}.
After the passage of $\phi \sim 0$, the $\chi$ mass increases and correspondingly the decay rate of $\chi$ also becomes large.
Assuming the typical $\chi$ decay rate as $\Gamma_\chi \sim \alpha m_\chi = \alpha \lambda |\phi(t)|$,
it decays at $t_{\rm dec}\sim (\alpha\lambda m_\phi\tilde\phi)^{-1/2}$,
well before the $\phi$ again reaches the maximum at the opposite side of the potential, if $m_\phi \ll \alpha\lambda\tilde\phi$.
Otherwise, $\phi$ returns back to $\phi=0$ before $\chi$ decays into radiation.
Then the parametric resonant amplification of the $\chi$ modes occurs for a bosonic $\chi$.
For a fermionic $\chi$, the Pauli blocking suppresses the further particle production.
Note that $\chi$ also has a thermal dissipation rate of order of $\Gamma_\chi \sim \alpha T$ for $\lambda\phi \lesssim T$,
but this does not dominate the above estimate as long as the condition (\ref{phimax}) is met.

%Above arguments neglect the interaction of $\chi$ with thermal plasma.
%If $\chi$ interacts frequently with particles in thermal bath, the resonant effect does not happen~\cite{Kasuya:1996aq}.
%A simple criterion is that thermal mass for the $\chi$, $m_\chi \sim gT$, must be smaller than $k_*$
%for efficient preheating.\footnote{
%	The non-adiabatic zone, $\Delta\phi_* \simeq \sqrt{m_\phi\phi_0/\lambda}$, 
%	is smaller than $\phi_c (= T/\lambda)$ for $T \gtrsim \sqrt{\lambda m_\phi \phi_0}$.
%}
%This condition is rewritten as $T \lesssim \sqrt{\lambda m_\phi\phi_0/(4\pi\alpha)}$.

(a)
In the case of $\lambda\tilde\phi \gg m_\phi/\alpha$, a phenomenon similar to the so-called instant preheating~\cite{Felder:1998vq}
takes place.
In each oscillation, the $\chi$ particle number density $(\ref{nchi})$ is produced, which soon decays into radiation.
Therefore, the fractional energy density which $\phi$ loses in one oscillation is given by
\begin{equation}
	\delta_\phi \equiv \frac{\delta \rho_\phi}{\rho_\phi} \simeq \frac{\lambda^2}{4\pi^3\sqrt{\alpha}}.
\end{equation}
As long as the preheating lasts, this fractional amount of energy dissipates into thermal bath in one oscillation.
The resultant $\phi$ energy density is given by
\begin{equation}
	\rho_\phi(t)\sim \rho_{\phi_i}(1-\delta_\phi)^{m_\phi t/ \pi}=\frac{1}{2}m_\phi^2\tilde\phi^2(1-\delta_\phi)^{m_\phi t/ \pi} .
\end{equation}
Unless $\delta_\phi$ is exceedingly small, the $\phi$ energy density efficiently is transformed into the radiation
and the amplitude $\tilde\phi$ decreases correspondingly.
The effective dissipation rate of $\phi$ is then given by\footnote{
	To be exact, the $\chi$'s degrees of freedom should be multiplied, but 
	we will not care about factors in what follows.
}
\begin{equation}
	\Gamma_\phi \sim \frac{1}{\pi }\delta_\phi m_\phi 
	= \frac{\lambda^2 m_\phi}{4\pi^4 \sqrt{\alpha}}.
\end{equation}
Notice that this dissipation exists in the regime the perturbative decay of $\phi$ into $\chi$ is prohibited, i.e.,
when $m^\chi_{\rm eff}{^2} \sim \lambda^2 \tilde\phi^2 + g^2T^2 \gg m_\phi^2$. 
This regime ends when the amplitude decreases and either of the following condition is satisfied.
(i) $\lambda\tilde\phi= 4\pi\alpha T^2/m_\phi$, where thermal mass of $\chi$ becomes efficient.
Then the particle production events end.
(ii) $\lambda\tilde\phi=m_\phi/\alpha$, where the regime of instant preheating ends.
Then it enters the regime (b) in the following.

(b)
In the case of $\lambda\tilde\phi \ll m_\phi/\alpha$, the produced particles around $\phi\sim 0$,
which fills the phase space density of $k\lesssim k_*$, survive until the $\phi$ again returns back to $\phi\sim 0$.
For a boson $\tilde\chi$, the efficiency of particle production increases due to the parametric resonance effect.
As a result, the $\tilde\chi$ number density exponentially increases until the backreaction terminates the resonance.
It is expected that at this stage the $\phi$ and $\chi$ energy density are equilibrated.
For a fermion $\chi$, on the other hand, no such resonant effect takes place due to the Pauli blocking effect.
In either case, at most $\mathcal O(1)$ fraction of the $\phi$ energy density is dissipated.

%%%%%%%%%%%%%%%%%%%%%%%%%%%%%
\subsection{The case of finite temperature mass} 
\label{sec:}
%%%%%%%%%%%%%%%%%%%%%%%%%%%%%

Next, let us consider the case where the $\phi$ 
oscillates by the finite-temperature effect,
and see whether or not the non-perturbative particle production occurs 
when the $\phi$ passes through the origin.
As already mentioned in~\ref{foot:inflaton},
we assume that the thermal plasma is already produced by the 
decay of inflaton before the scalar condensate in consideration starts to
oscillate.\footnote{
	At the onset of oscillation,
	typical thermalization time scale of the plasma is $(\alpha T)^{-1}$, which is much faster than the
	oscillation time scale of the scalar field for our parameter choices in the following section, hence our assumption is justified.
	If this is not the case, the formation of thermal plasma and its effects on scalar dynamics require more careful treatment.
	See e.g. Ref.~\cite{Allahverdi:2011aj,Allahverdi:2010xz} for further discussion on the issue of thermalization after inflation.
}

First note that the rate of thermalization in thermal bath, $\Gamma_{\rm th} \sim \alpha T$,
is larger than the effective mass scale of $\phi$ at the beginning of oscillation,
$\Gamma_{\rm th} \gg H \sim m_{\rm eff}^\phi$.
%Thus, in general, we can safely apply the study given in Sec.\ \ref{sec:thermal}
%at the first several oscillations.
%except for the region where the adiabaticity of the coupled particles $\chi$ is broken down.
This observation immediately means that 
if the $\phi$ begins to oscillate with the thermal mass potential,
then the motion of $\phi$ is %always 
adiabatic with respect to the thermalization time scale of
thermal plasma, $m_{\rm th}^\phi (T) \sim \lambda T \ll \Gamma_{\rm th}\sim \alpha T$,
as far as it oscillates with the thermal mass.
%Hence, this situation results in the case (i) discussed below.
Hence, the discussion given in Sec.\ \ref{sec:small} is applicable at all times
as shown below the case (i).

Second, if the $\phi$ oscillates with the thermal logarithmic potential 
(not necessarily at the beginning of oscillation),
then the typical thermalization time scale of thermal plasma is larger than
the effective mass scale of $\phi$, $m_{\rm eff}^\phi \sim \alpha T^2/\phi$,
because of $\phi > \phi_c = T/\lambda$.
Thus the discussion given in Sec.\ \ref{sec:large} is applicable in this case,
and we can reliably use the thermal logarithmic potential for $\phi_c < \phi (t)$.
At $\phi (t) < \phi_c$, the $\chi$ particles begin to be produced from thermal plasma and 
they yield a finite density correction to $\phi$'s equation of motion.

At this stage, we have to distinguish two cases.
\\

(i) $\lambda \ll \alpha$\\
In this case, the motion of  $\phi$ is adiabatic with respect to the relaxation 
time scale of thermal plasma. Thus, the discussion given in Sec.\ \ref{sec:small} is 
applicable. As discussed in Sec.\ \ref{sec:small}, the finite density correction 
can be expressed as
\begin{align}
	\int d\tau\ \Pi_{\rm ret} (t-\tau) \phi (\tau) 
	\simeq \
	m_{\rm th}^\phi (T)^2 \phi + \Gamma_\phi \dot{\phi}
\end{align}
where the thermal mass is given by $m_{\rm th}^\phi (T) \sim \lambda T$.
Then, let us see whether the non-perturbative particle production occurs or not.
The adiabaticity parameter is estimated to be
\begin{align}
	\left| \frac{\dot{\omega_\chi}}{\omega_\chi^2} \right|
	<
	\frac{\lambda^2 \phi_c}{\alpha T}
	=
	\frac{\lambda}{\alpha}
	\ll 1
\end{align}
where $\omega_\chi^2 = k^2 + m_{\rm eff}^\chi{^2}$.
In the first equality, we have substituted $k=0$
and used the fact that the potential energy of thermal logarithmic potential is small 
compared to one of thermal mass potential, $\alpha^2 T^4 \ll T^4$.
In the second equality, we have used $\phi_c = T/\lambda $. This shows that 
no non-perturbative particle production occurs while the $\phi$ oscillates with the thermal mass
for $\lambda \ll \alpha$.

Therefore,  for $\lambda \ll \alpha$, the thermal potential derived in Sec.\ \ref{sec:thermal}
is applicable at all times.\\

(ii) $\lambda \gg \alpha$\\
In this case, the $\phi$ oscillation frequency is much larger than the thermalization rate.
Hence, the propagators of $\chi$ cannot be considered as thermal ones and
have to be treated as dynamical ones.
There are two dominant effects that produce the $\chi$ particles:
production from thermal plasma and non-perturbative production. 
The essential difference from the case (i) is that even if the produced $\chi$ particles become
heavy due to the field value of $\phi$, the produced particles may not decay and survive
at each oscillation of $\phi$, depending on the balance between 
the oscillation time scale of $\phi$ and the decay rate of $\chi$.
The survived heavy particles significantly affect the motion of $\phi$, since they
form a linear potential effectively:
\begin{align}
	M^2 \phi 
	\sim
	\lambda^2 \int \frac{d^3 k}{(2\pi)^3} \frac{f_\chi (\omega_{\bf k})}{\omega_{\bf k}} \ \phi
	\sim  
	\lambda n_{\chi},
\end{align}
as in the case with the moduli trapping~\cite{Kofman:2004yc}.
The motion of $\phi$ is determined by the complicated balance of parameters,
and so, in the following, we will not consider the case (ii) for simplicity.
The complete analysis of the trapping effect in the presence of thermal plasma 
will be performed elsewhere.\\

%%%%%%%%%%%%%%%%%%%%%%%%%%%%%
\section{Dynamics of oscillating scalar field} 
\label{sec:dyn}
%%%%%%%%%%%%%%%%%%%%%%%%%%%%%

Basic ingredients for analyzing the scalar dynamics have been presented above.
Let us follow the evolution of the scalar field.
First, we divide three cases depending on which term in the effective potential determines the oscillation epoch of $\phi$:
zero-temperature mass, thermal mass or thermal log.

\begin{itemize}

\item 
Case (a): $\phi$ begins to oscillate with thermal logarithmic potential if
\begin{equation}
	\phi_i < \phi_i^c\equiv \alpha T_{\rm R}\sqrt{\frac{M_P}{m_\phi}}
	~~~{\rm and}~~~\lambda \phi_i^{3/2} > T_{\rm R}(\alpha M_P)^{1/2}.
\end{equation}

\item 
Case (b): $\phi$ begins to oscillate with thermal mass if
\begin{equation}
	\lambda > \lambda^c\equiv \left( \frac{m_\phi^3}{T_{\rm R}^2 M_P} \right)^{1/4}
	~~~{\rm and}~~~\lambda \phi_i^{3/2} < T_{\rm R}(\alpha M_P)^{1/2}.
\end{equation}

\item 
Case (c): Otherwise, $\phi$ begins to oscillate with zero-temperature mass.

\end{itemize}

Note that we assume that $\chi$ particles are absent initially in the case (a).
Otherwise, $\phi$ would feel correction to the effective potential.
This assumption might break down if heavy $\chi$ particles are substantially produced by the direct decay of inflaton
or the inflaton preheating process.
Whether this occurs or not depends on the inflation model, and hence we simply assume the absence of $\chi$ in the case (a).

\begin{figure}
\begin{center}
\includegraphics[scale=1.3]{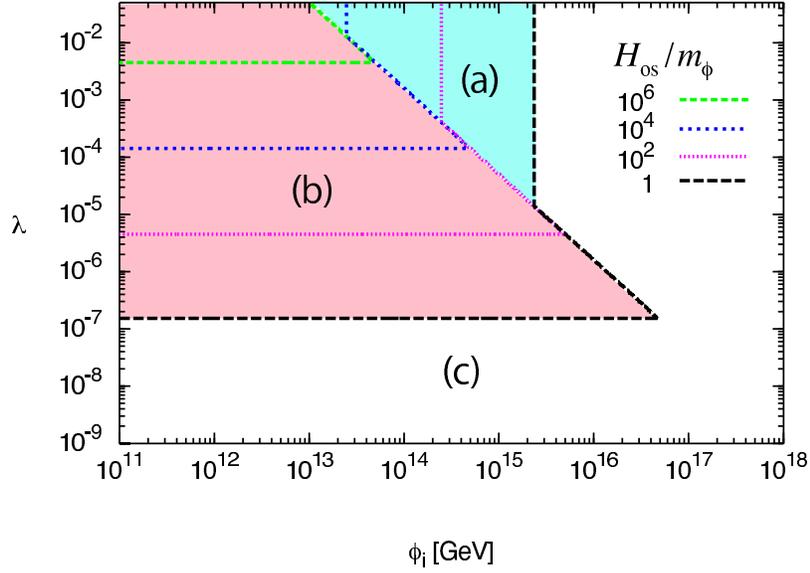}
\vskip 1cm
\includegraphics[scale=1.3]{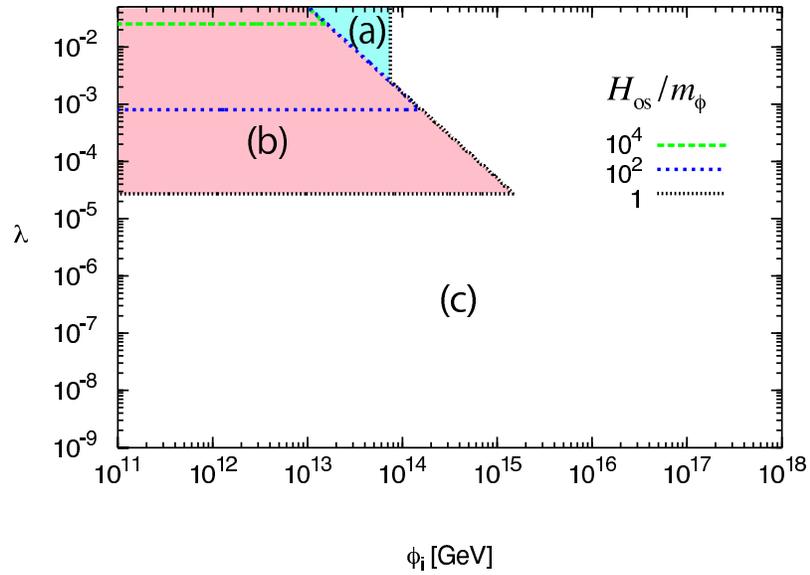}
\caption{ (Top) Contours of $H_{\rm os}/m_\phi$ for $\alpha=0.05, T_{\rm R}=10^9$\,GeV and $m_\phi=1$\,TeV.
(Bottom) Same as top panel, but for $m_\phi=10^3$\,TeV. }
\label{fig:Hos}
\end{center}
\end{figure}

We take parameters so that $T_{\rm R}^2/M_P < m_\phi$ is satisfied in all the following analysis.
Therefore, in any case, $\phi$ begins to oscillate during the inflaton oscillation dominated phase.
The Hubble parameter at the beginning of oscillation, $H_{\rm os}$, is then given by
\begin{equation}
	H_{\rm os} \simeq 
	\begin{cases}
		\alpha^2 T_{\rm R}^2 M_P/\phi_i^2&~{\rm for ~case~(a)},\\
		\left(\lambda^4 T_{\rm R}^2 M_P \right)^{1/3} &~{\rm for ~case~(b)},\\
		m_\phi &~{\rm for ~case~(c)}.
	\end{cases}
\end{equation}
Fig.~\ref{fig:Hos} shows contours of $H_{\rm os}/m_\phi$ on $(\phi_i, \lambda)$-plane 
for $\alpha=0.05, T_{\rm R}=10^9$\,GeV and $m_\phi=1$\,TeV (top) and for $m_\phi=10^3$\,TeV (bottom).
Regions (a) -- (c) correspond to the cases (a) -- (c) described above.
Note that it is possible that even if $\phi$ begins to oscillate with zero-temperature mass term,
thermal effects become dominant thereafter.
Conversely, the zero-temperature term eventually becomes dominant even if thermal effects are important at the
beginning of oscillation.
These facts make the scalar dynamics quite complicated.

%%%%%%%%%%%%%%%
%%%%%%%%%%%%%%%

The evaluation of decay or evaporation epoch of $\phi$ is even more complicated
due to the kinematical condition including the field-dependent and temperature-dependent mass of $\chi$
as well as the existence of thermal dissipation rate.
In order to follow the dynamics of oscillating scalar field,
it is convenient to study averaged quantities with a time interval which is longer than
the oscillation period but shorter than the Hubble time scale.
We summarize useful equations in Appendix\ \ref{sec:ev_osc_scalar}.
The averaged effective dissipation rate,
$\Gamma_\phi^{\rm eff}$, at an each averaged amplitude $\tilde\phi$ regime,
which is defined in Appendix\ \ref{sec:ev_osc_scalar},
is summarized as follows.

\begin{itemize}

\item 
If $\phi$ oscillates with thermal log potential $(m^\phi_{\rm eff}  \sim\alpha T^2/\tilde\phi)$, 
the dissipation is caused by scattering with gauge bosons in thermal bath.
Notice that the $\phi$ decay into gauge boson pair is kinematically forbidden.
The dissipation rate is evaluated as
\begin{equation}
	\Gamma_\phi^{\rm eff} \sim \frac{b\, \alpha^2 T^3}{\tilde \phi \phi_c}.
\end{equation}
We have shown that there are no efficient non-perturbative production in this regime for $\lambda \lesssim \alpha$.

\item 
If $\phi$ oscillates with a thermal mass term $(m^\phi_{\rm eff} \sim\lambda T)$, 
we need to distinguish two cases: $\lambda < \alpha$ and $\lambda > \alpha$.
In the former case, the oscillation is adiabatic with respect to thermal relaxation rate
and the dissipation is caused by scattering with $\chi$ particles in thermal bath.
In the latter case, $\phi$ can kinematically decay into $\chi$ pair. However, the dynamics is so so complicated
and we do not consider this case. Thus we have
\begin{equation}
	\Gamma_\phi^{\rm eff} \sim
		\lambda^2 \alpha T  {\rm ~~~for~~~} %m^\phi_{\rm eff} < \alpha T
		\lambda \tilde\phi \ll \alpha T, 
\end{equation}
As a rough approximation, we simply extrapolate this result to 
$\alpha T < \lambda \tilde\phi \ll T$.\footnote{
	For numerical simplicity,
	we do not use the dissipation coefficient computed from Eq.\ \eqref{eq:spec_fer_dirac}
	which is applicable at $\alpha T < \lambda \tilde\phi \ll T$.
	\label{fn:dissipation}
}
We have shown that there are no efficient non-perturbative production in this regime.

\item 
If $\phi$ oscillates with a zero-temperature mass term $(m^\phi_{\rm eff} \sim m_\phi)$, we have several situations.
First, consider the case of $m_{\rm eff}^\phi \ll \alpha T$.
%$\lambda\tilde \phi \gg T$.
%In this case $\chi$ decouples from thermal bath. The dissipation rate is given by
In this case, the dissipation coefficient can be evaluated similarly, since
the motion of $\phi$ is adiabatic with respect to the thermalization time scale of hot plasma.
\begin{align}
	\Gamma_\phi^{\rm eff} \sim
	\begin{cases}
		b\, \alpha^2 T^3/(\tilde\phi  \phi_{\rm th}) & {\rm ~for~~} T \ll \lambda \tilde\phi, \\
		\lambda^2 \alpha T  & {\rm ~for~~} \lambda \tilde\phi \ll \alpha T. \\
	\end{cases}
\end{align}
As a rough approximation, we extrapolate these results to $\alpha T < \lambda \tilde\phi < T$.\footnote{
	See footnote \ref{fn:dissipation}.
}
Then, the threshold $\phi_{\rm th}$ can be estimated as 
$gT/\lambda$ if the non-perturbative production is absent
or $\sqrt{m_\phi \tilde\phi / \lambda}$,
where the adiabaticity is broken,
 if it is present.
Next, consider the case of $m_{\rm eff}^\phi \gg gT$.
In this case, $\Gamma_\phi^{\rm eff}$ can be estimated with neglecting the finite density correction to
the dispersion relation of $\chi$.\footnote{
	As can be seen from Eqs.~\eqref{eq:small_dis_f} and \eqref{eq:small_dis_b}, 
	there are Pauli blocking or Bose enhancement
	factors, though we will neglect them for simplicity in the following.
}

%$\lambda\tilde \phi \ll T$.
%The dissipation rate depends on whether the decay into $\chi$ pair is kinematically allowed or not.
%It is evaluated as
%%
\begin{equation}
	\Gamma_\phi^{\rm eff} \sim 
	\begin{cases}
		\alpha^2 m_\phi{^3}/ (\tilde\phi  \phi_{\rm th})
		& {\rm ~for~~} m_\phi \ll  \lambda \tilde\phi, \\
		\lambda^2 m_\phi/ (8\pi)  & {\rm ~for~~} \lambda \tilde\phi \ll m_\phi.    
	\end{cases}
\end{equation}
The threshold $\phi_{\rm th}$ is evaluated as $\sqrt{m_\phi \tilde\phi/ \lambda}$.
Though we do not calculate $\Gamma_\phi$ at 
the regime $\alpha T < m^\phi_{\rm eff} < gT$ actually,
we simply extrapolate these results as approximate ones.
As pointed out in Sec.~\ref{sec:non-zero}, the non-perturbative particle production gives the effective dissipation rate as
\begin{equation}
	\Gamma_\phi \sim \frac{\lambda^2 m_\phi}{4\pi^4 \sqrt\alpha} {\rm~~~for~~~}
	\lambda \tilde \phi \gg {\rm max} \left\{ m_\phi, \frac{g^2T^2}{m_\phi} \right\}
\end{equation}

\end{itemize}

\begin{figure}
\begin{center}
\includegraphics[scale=1.5]{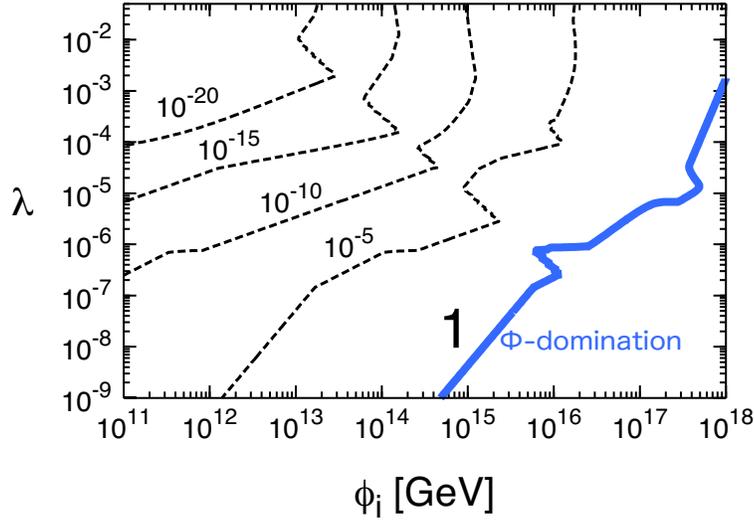}
\vskip 1cm
\includegraphics[scale=1.5]{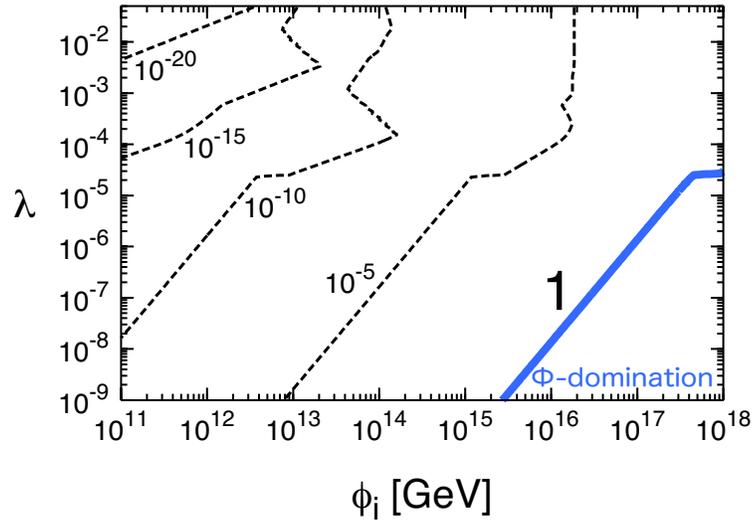}
\caption{(Top) Contours of $R_\phi$ for $\alpha=0.05, T_{\rm R}=10^9$\,GeV and $m_\phi=1$\,TeV.
(Bottom) Same as top panel, but for $m_\phi=10^3$\,TeV. 
In the region labeled by ``{\color{blue}$\phi$-domination}'', the $\phi$ energy density dominates the Universe before it decays. }
\label{fig:R}
\end{center}
\end{figure}

Given above formulae, we can trace the evolution of scalar field oscillation.
One more complexity arises from the fact that the time dependence of quantities $T$, $\tilde \phi$ and so on
changes before and after the reheating.
After the reheating, $\phi$ can again dominate the Universe depending on parameters.
The evolution equations of the system are given by
\begin{align}
	&\ddot\phi + (3H+\Gamma_\phi (\phi;T))\dot\phi + 
	\frac{\der V(\phi;T)}{\der \phi} = 0, \\
	&\dot\rho_r +4H\rho_r= \Gamma_{\rm inf}\rho_{\rm inf} + \Gamma_\phi \rho_\phi, 
	\label{eq:rad}\\
	&H^2 = \cfrac{1}{3M_P^2}(\rho_{\rm inf}+\rho_\phi+\rho_r),
\end{align}
where $V := m_\phi^2 \phi^2 / 2 + V_{\rm eff}$, 
$\rho_r = (\pi^2 g_*/30)T^4$ is the radiation energy density at the leading order 
in $g$ and $\lambda$,
%$\rho_\phi = m_\phi^2 \tilde \phi^2 /2$
and $\rho_{\rm inf}$ is the inflaton energy density and $\Gamma_{\rm inf} = \sqrt{(\pi^2 g_*/90)}T_{\rm R}^2/M_P$ 
is the inflaton decay rate. 
The energy density of $\phi$\footnote{
	The energy density should not be confused with the free energy density.
} is given by
\begin{align}
	\rho_\phi = \left< \overline{\frac{1}{2} \dot\phi^2 + \frac{1}{2} m_\phi^2 \phi^2} \right>
	= \left< \overline{\phi \frac{\der V}{\der \phi}} \right> + \frac{1}{2} m_\phi^2 \tilde \phi^2.
\end{align}
In the second equality, 
we have used the virial theorem (See also Appendix\ \ref{sec:ev_osc_scalar}). 
Here $\< \overline{\cdots} \>$ denotes the time average with a time interval which is longer than
the oscillation period but shorter than the Hubble time scale.
Validity of the last term in Eq.\ \eqref{eq:rad} is restricted to the case when the $\phi$
oscillates with the zero temperature mass term, however,
it is sufficient for practical use because the dominant change of temperature due to the
loss of $\phi$'s energy occurs when the $\phi$'s energy dominate the universe,
that is, when the $\phi$ oscillates with the zero temperature mass term.
Following scaling relations will be useful 
(See Appendix~\ref{sec:ev_osc_scalar}):
\begin{equation}
	\tilde\phi \propto 
	\begin{cases}
		a^{-3/2}           & {\rm ~for~~}\mbox{zero temperature mass}, \\
		a^{-3/2}T^{-1/2} & {\rm ~for~~}\mbox{thermal mass},\\
		a^{-3}T^{-2} & {\rm ~for~~} \mbox{thermal log}.\\  
	\end{cases}
\end{equation}
The scalar field $\phi$ is expected to evaporate when $H=\Gamma_\phi (T,\tilde\phi)$ is satisfied.
In order to see how the $\phi$'s amplitude evolves
and to clarify thermal effects on the evolution of $\phi$'s amplitude, it is convenient to consider the quantity
\begin{equation}
	R_\phi \equiv \left.\frac{ m_\phi^2 \tilde \phi^2 / 2}{\rho_{\rm inf}+\rho_r}\right| _{H=\Gamma_\phi}.
\end{equation}
In Fig.~\ref{fig:R} we have plotted contours of $R_\phi$ for $\alpha=0.05, T_{\rm R}=10^9$\,GeV and $m_\phi=1$\,TeV
(top) and $m_\phi=10^3$\,TeV (bottom).
In the region labeled by ``$\phi$-domination'', the $\phi$ energy density dominates the Universe before it decays.
It is seen that the behavior is highly non-trivial.
This kind of complicated structure would not appear without taking thermal effects into account.
In order to see what is happening, let us see some typical cases in these parameter regions.

%%%%%%%%%%%%%%%
%%%%%%%%%%%%%%%

%%%%%%%%%%%%%%%%%%%%%%%%%%%%%
\subsection{Oscillation with thermal log} 
%%%%%%%%%%%%%%%%%%%%%%%%%%%%%

After $\phi$ begins to oscillate with thermal logarithmic potential, the amplitude of $\phi$,
decreases as $\tilde \phi \propto a^{-9/4}$ while the temperature decreases as $T\propto a^{-3/8}$.
As described in Sec.~\ref{sec:large}, in this regime $\chi$ has a large mass and is decoupled from thermal bath.
By integrating out the heavy $\chi$ field, $\phi$ has an effective interaction with gauge fields
as $\mathcal L \sim \delta\phi FF / \tilde \phi$.
Note that the effective mass of $\phi$ may be estimated as 
$m^\phi_{\rm eff} \sim \alpha T^2 /\tilde \phi \ll gT$,
hence the decay into gauge bosons are kinematically forbidden.
However, $\phi$ receives dissipative effects from thermal bath
and the dissipation rate is given by 
$\Gamma_\phi^{\rm eff} \sim b \alpha^2 T^3/(\tilde \phi \phi_c)$.

\begin{figure}
\begin{center}
\includegraphics[scale=1.3]{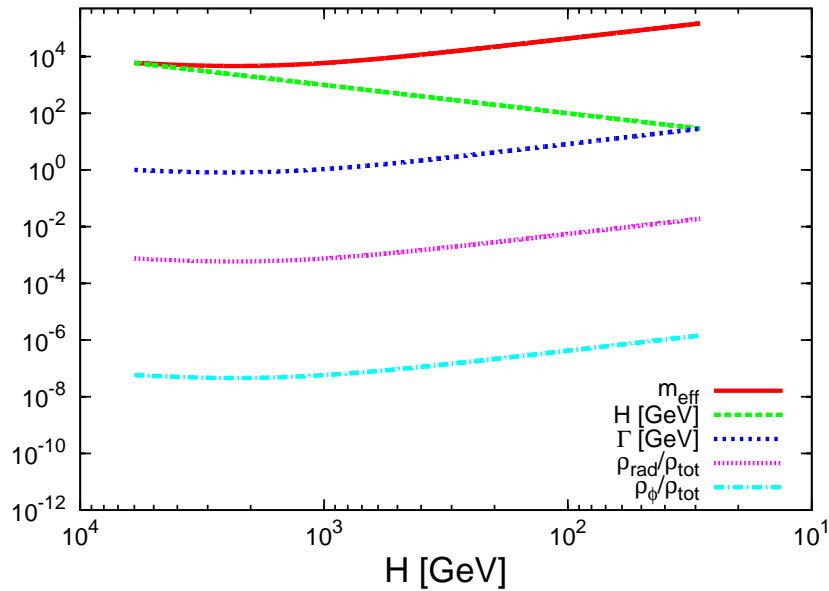}
\caption{ 
Time evolution of various quantities as a function of Hubble scale for
$(\lambda, \phi_i) = (2 \times 10^{-3},10^{15}\,{\rm GeV})$. We have taken $m_\phi = 1$\,TeV and $T_{\rm R}=10^9$\,GeV.}
\label{fig:Tlog}
\end{center}
\end{figure}

Fig.~\ref{fig:Tlog} shows time evolution of various quantities as a function of Hubble scale.
At $H=H_{\rm os} \sim 6\times 10^3$\,GeV, $\phi$ begins to oscillate with thermal logarithmic potential.
As the amplitude decreases, the dissipation rate increases
$\Gamma_\phi^{\rm eff}\propto a^{3/2} \propto H^{-1}$
and it becomes equal to the Hubble parameter $H$ at $H_{\rm dec} \sim 3 \times 10$\,GeV,
where $\phi$ is expected to evaporate.
In this case, therefore, the $\phi$ coherent oscillation soon disappears after the onset of oscillation
due to the interaction with thermal plasma.\footnote{
	It may be the case that $\phi$ coherent oscillation deforms into non-topological solitons, oscillons,
	before the evaporation. Since the time scale of the development of spatial instability
	is comparable to the evaporation time scale, we avoid definite conclusion about this issue.
	The case of oscillon formation will be discussed in the Appendix~\ref{sec:non_top}.
}

%%%%%%%%%%%%%%%
%%%%%%%%%%%%%%%

%%%%%%%%%%%%%%%%%%%%%%%%%%%%%
\subsection{Oscillation with thermal mass} 
%%%%%%%%%%%%%%%%%%%%%%%%%%%%%

\begin{figure}
\begin{center}
\includegraphics[scale=1.3]{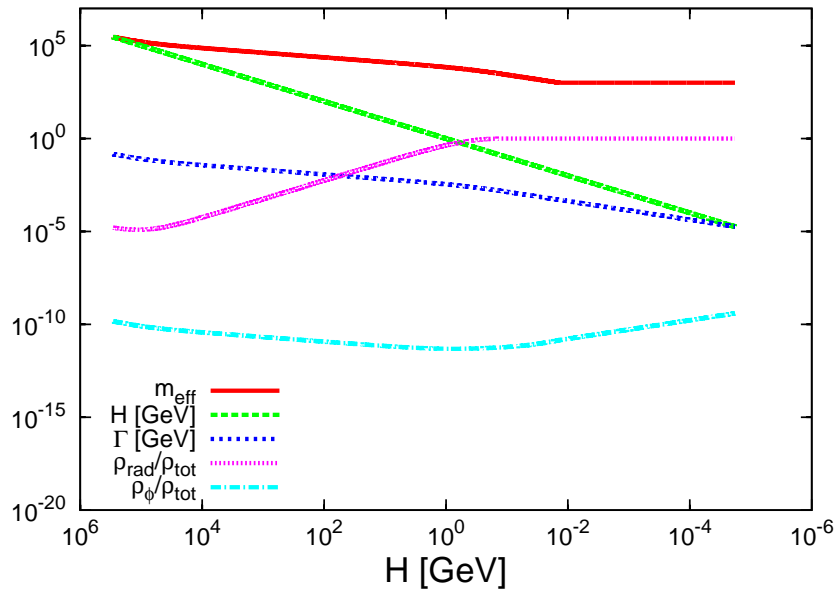}
\caption{ 
Same as Fig.~\ref{fig:Tlog}, but for
$(\lambda, \phi_i) = (10^{-5},10^{14}\,{\rm GeV})$. }
\label{fig:Tmass}
\end{center}
\end{figure}

Next, we consider a typical case where thermal mass plays an important role.
Fig.~\ref{fig:Tmass} shows time evolutions of various quantities as a function of Hubble scale for
$(\lambda, \phi_i) = (10^{-5},10^{14}\,{\rm GeV})$, $m_\phi = 1$\,TeV and $T_{\rm R}=10^9$\,GeV.
As can be seen in Fig.~\ref{fig:Hos}, $\phi$ begins to oscillate with a thermal mass for this parameter choice
at $H=H_{\rm os} \sim 10^6$\,GeV.
As the temperature decreases, thermal mass of $\phi$ also decreases as
$m_{\rm eff}^\phi = \lambda T \propto H^{1/4}$ in the matter dominated era and $m_{\rm eff}^\phi\propto H^{1/2}$
in the radiation dominated era after the reheating.
At the temperature $T\sim m_\phi/\lambda = 10^8$\,GeV, or $H\sim 10^{-2}$\,GeV,
the zero-temperature mass begins to dominate.
Since $m_\phi \ll \alpha T$ at this stage, the motion of $\phi$ is adiabatic and 
the main dissipation effect comes from the scattering with $\chi$ in thermal bath.
The dissipation coefficient is given by $\Gamma_\phi \simeq \lambda^2\alpha T$.
Thus $\phi$ evaporates at $H=H_{\rm dec} \sim 10^{-5}$\,GeV, where
\begin{equation}
	H_{\rm dec} \sim \lambda^4 \alpha^2 M_P.
\end{equation}
The temperature at the evaporation is then estimated to be $T_{\rm dec} \sim 10^7$\,GeV,
consistent with the assumption that it occurs after the reheating: $T_{\rm dec} < T_{\rm R}$.

%%%%%%%%%%%%%%%
%%%%%%%%%%%%%%%

%%%%%%%%%%%%%%%%%%%%%%%%%%%%%
\subsection{Oscillation with zero-temperature mass} 
%%%%%%%%%%%%%%%%%%%%%%%%%%%%%

\begin{figure}
\begin{center}
\includegraphics[scale=1]{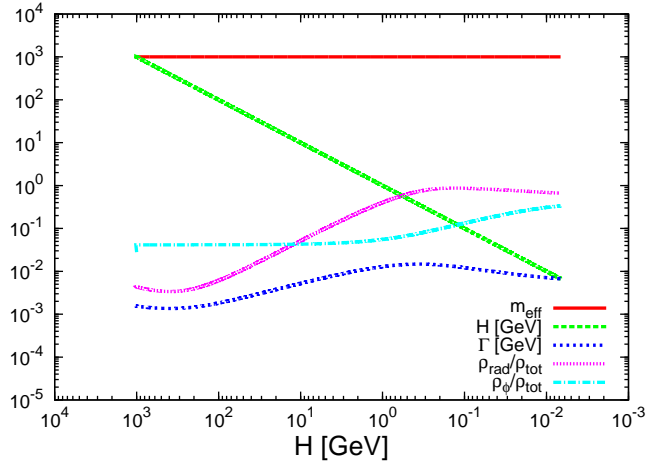}
\vskip 0.5cm
\includegraphics[scale=1]{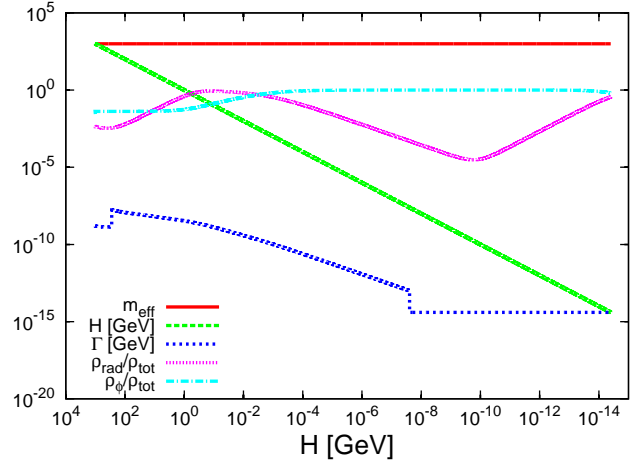}
\vskip 0.5cm
\includegraphics[scale=1]{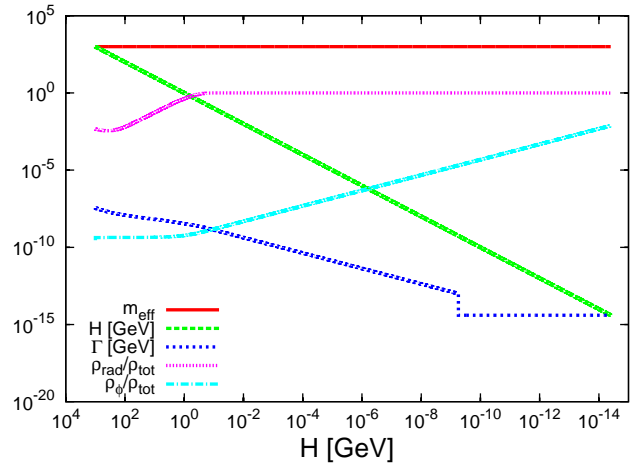}
\caption{Same as Fig.~\ref{fig:Tlog}, but for $(\lambda, \phi_i) = (10^{-2},10^{18}\,{\rm GeV})$ (top), 
  $ (10^{-8},10^{18}\,{\rm GeV})$ (middle),  $(10^{-8},10^{14}\,{\rm GeV})$ (bottom). }
\label{fig:zeromass}
\end{center}
\end{figure}

Let us consider the case where $\phi$ begins to oscillate with zero-temperature mass.
In the top panel of Fig.~\ref{fig:zeromass}, we show 
time evolutions of various quantities as a function of Hubble scale for
$(\lambda, \phi_i) = (10^{-2},10^{18}\,{\rm GeV})$, $m_\phi = 1$\,TeV and $T_{\rm R}=10^9$\,GeV.
After the reheating, $\phi$ soon dominates the Universe at $H\sim 10^{-3}$\,GeV.
Since the amplitude is so large, $\chi$ obtains a large mass and decouples from thermal bath.
By noting that $\phi$ decay into gauge bosons is kinematically forbidden due to large thermal mass,
the main dissipative effect comes from the scattering with gauge bosons in thermal bath.
The dissipative coefficient is given by $\Gamma_\phi \sim b \alpha^2T^3/(\tilde\phi \phi_{\rm th})$.
These thermal particles scatter off $\phi$ and as a result, $\phi$ evaporates at $H\sim 10^{-5}$\,GeV.

In the middle panel of Fig.~\ref{fig:zeromass}, we show 
time evolutions of various quantities as a function of Hubble scale for
$(\lambda, \phi_i) = (10^{-8},10^{18}\,{\rm GeV})$, $m_\phi = 1$\,TeV and $T_{\rm R}=10^9$\,GeV.
In this case, $\lambda$ is so small that the dissipation rate is suppressed compared with the above case.
After the Hubble parameter decreases to $H\sim 10^{-8}$\,GeV, 
the amplitude of $\phi$ becomes small so that the $\phi$ decay into $\chi$ pair is accessible.
Finally radiation generated from $\phi$ dominates over the inflaton decay products.
The decay rate at this stage is given by $\Gamma_\phi \sim \lambda^2 m_\phi/(8\pi)$.
Therefore, $\phi$ decays at $H=H_{\rm dec} \sim \lambda^2 m_\phi/(8\pi) \sim 10^{-15}$\,GeV.

In the bottom panel of Fig.~\ref{fig:zeromass}, we show 
time evolutions of various quantities as a function of Hubble scale for
$(\lambda, \phi_i) = (10^{-8},10^{14}\,{\rm GeV})$, $m_\phi = 1$\,TeV and $T_{\rm R}=10^9$\,GeV.
The situation is similar to the former case, but in this case the amplitude of $\phi$ is so small
and $\phi$ never dominates the Universe.
Thus $\phi$ decays at $H=H_{\rm dec} \sim \lambda^2 m_\phi/(8\pi) \sim 10^{-15}$\,GeV,
when the cosmic temperature is around $T\sim 100$\,GeV.

%%%%%%%%%%%%%%%
%%%%%%%%%%%%%%%

%%%%%%%%%%%%%%%%%%%%%%%%%%%%%
\section{Conclusions and discussion} 
\label{sec:conc}
%%%%%%%%%%%%%%%%%%%%%%%%%%%%%

In this paper we have extensively studied the scalar dynamics in the early Universe
taking account of the effects of thermal environment.
Despite the fact that the early Universe after inflation may be filled with high-temperature thermal plasma,
thorough analyses on the scalar dynamics in such an environment were missing in the literature.
Needless to say, scalar fields play important roles in cosmology.
Inflaton explains the primordial inflation and the primordial density perturbation, 
and curvaton may also be responsible for the generation of density perturbation.
Affleck-Dine fields in the MSSM can create the baryon asymmetry of the Universe.
SUSY breaking fields may have significant effects on cosmology.
Similar long-lived scalars often appear in extensions of the SM.
All these scalar fields generally have large energy densities so that they must decay or evaporate
at an appropriate epoch in order not to disturb the success of standard cosmology.
Therefore, they necessarily couple to SM particles directly or indirectly through some intermediate states.

Based on these observations, we considered a model in which a scalar $\phi$ couples to 
fermions $\chi$ through a Yukawa coupling, which then interact with thermal bath.
This simple model captures essential features of realistic models.
We have consistently taken into account various effects:
thermal modification on the effective potential of $\phi$, thermal dissipation of $\phi$,
non-perturbative particle production and the formation of non-topological solitons.
It is found that even in this simple class of models, the scalar dynamics is so complicated
that a naive estimation neglecting these effects is not allowed in broad parameter spaces.
In particular, it should be noticed that thermal dissipation, which can be interpreted as the $\phi$ evaporation
due to scatterings with particles in an environment, often plays a dominant role in 
determining the dissipation epoch of $\phi$.

Finally, let us mention what we have not included in our analyses.
In this paper we have followed the evolution of only the zero-mode of $\phi$ coherent oscillation.
However, it is expected that fluctuations around the zero-mode, or the particle-like excitations of $\phi$
can be important if thermal effects play a dominant role for the evaporation of $\phi$.
This can in principle be traced by following the evolution of the two-point functions of $\phi$.
We have also restricted ourselves to the case of $\lambda < \alpha$, because otherwise 
the whole dynamics including $\chi$ particle production is complicated. 
We will return to these issues in some concrete models elsewhere~\cite{mukaida}.

%%%%%%%%%%%%%%%%%%%%%%%%%%%%%%%%%%%%
\section*{Acknowledgment}
%%%%%%%%%%%%%%%%%%%%%%%%%%%%%%%%%%%%

This work is supported by Grant-in-Aid for Scientific research from
the Ministry of Education, Science, Sports, and Culture (MEXT), Japan,
No.\ 21111006 (K.N.) and No.\ 22244030 (K.N.) and also 
by World Premier International Research Center
Initiative (WPI Initiative), MEXT, Japan. 
The work of K.M. is supported in part by JSPS Research Fellowships
for Young Scientists.

%%%%%%%%%%%%%%%%%%%%%%%%%%%%%%%%%%%%
\appendix
%%%%%%%%%%%%%%%%%%%%%%%%%%%%%%%%%%%%

%%%%%%%%%%%%%%%%%%%%%%%%%%%%%
\section{Formation of non-topological solitons} 
\label{sec:non_top}
%%%%%%%%%%%%%%%%%%%%%%%%%%%%%

We have considered dynamics of oscillating scalar field with a quadratic mass term.
However, the mass receives radiative corrections from the fields 
that couple to $\phi$, and it may modify the mass like $m_\phi^2(\phi) \simeq m_\phi^2(1+\epsilon \ln(\phi))$,
where $\epsilon$ represents a one loop factor including coupling constants.
It is known that a coherently oscillating scalar field with a potential flatter than the 
quadratic potential ($\epsilon < 0$) may exhibit an instability that results in formations of classical lumps.
Moreover, thermal logarithmic potential may also cause such effects.\footnote{
	It is not clear the oscillon formation takes place in the case of thermal logarithmic potential.
	On the one hand, it is not known whether such a stable scalar configuration exists when the potential
	significantly deviates from the quadratic one, such as logarithmic form.
	On the other hand, even if such a solution exists,
	time scale required for the development of oscillon configuration is more or less close to the
	dissipation time scale in the case of thermal logarithmic potential.
	These facts make it difficult to estimate the probability for the oscillon formation.
}
These are called oscillons and studied in the context of inflaton fragmentation~\cite{Copeland:1995fq,McDonald:2001iv,Copeland:2002ku,Gleiser:2009ys,Amin:2010dc,Gleiser:2011xj}.
In Ref.~\cite{Kasuya:2002zs}, it is dubbed as ``I-ball'' in analogy with the Q-ball whose stability is ensured by a
conserved U(1) charge.
Similarly, the stability of I-ball is guaranteed by the existence adiabatic invariant in association with the scalar dynamics, 
which is often denoted by $I$.
Once the scalar field fragments into oscillons, their subsequent cosmological evolutions may be modified.
Although it may have non-trivial consequences, the dynamics is highly non-linear and
it is difficult to derive robust conclusions without numerical simulation.
Thus here we will only shortly see possible effects of oscillon formation in this Appendix.

%%%%%%%%%%%%%%%%%%%%%%%%%%%%%
\subsection{Oscillons} 
\label{sec:}
%%%%%%%%%%%%%%%%%%%%%%%%%%%%%

First we summarize the properties of the oscillon.
The most discussion follows from that performed in the context of Q-balls~\cite{Coleman:1985ki,Cohen:1986ct,Kusenko:1997si,Enqvist:1997si,Laine:1998rg,Banerjee:2000mb}.

As the $\phi$ begins to oscillate, low frequency modes exhibit instabilities to form the oscillon.
The most important mode is $k \sim H_{\rm os} \sim m_\phi$ 
$(k \sim H_{\rm os} \sim \alpha^2 T_{\rm R}^2M_P/\phi_i^2$ in the case of thermal log),
and the corresponding oscillon solution has a typical radius of $R\sim 1/H_{\rm os}$.
According to Ref.~\cite{Kasuya:2002zs}, the oscillon configuration corresponds to the bounce solution
with fixed adiabatic invariant, $I$.
In this case, it corresponds to the particle number density: $m_\phi^{\rm eff}\phi^2$.
Therefore, the total charge of an oscillon is estimated as
\begin{equation}
	I \simeq \beta\frac{4\pi R^3}{3}H_{\rm os}\phi_i^2 \sim \frac{4\pi \beta}{3}\left( \frac{\phi_i}{H_{\rm os}} \right)^2,
\end{equation}
where $\beta$ is a numerical factor which represents the delay of oscillon formation
from the epoch of $H=H_{\rm os}$.

Notice that even if $\phi$ begins to oscillate with zero-temperature mass term with positive $\epsilon$, 
thermal correction may come to dominate the effective potential depending on parameters.
In this case the ``delayed''-type oscillons may be formed~\cite{Kasuya:1999wu} if logarithmic potential allows the oscillon solution,
which we do not consider further. 
%Since the time scale of $\phi$ oscillation is much faster than the Hubble scale and the dissipation rate,
%the oscillon formation may be inevitable in this case.

%Fig.~\ref{fig:Iball} shows the parameter regions where
%oscillons may be formed for $m_\phi=1$\,TeV (black thick) and $m_\phi=10^3$\,TeV (blue thin).
%We have taken $T_{\rm R}=10^9$\,GeV. 
%Inside these lines, $\phi$ experiences the era of thermal log dominated regime before it evaporates 
%and hence oscillon formation potentially affects the fate of $\phi$.

%%%%%%%%%%%%%%%%%%%%%%%%%%%%%
\subsection{Decay of oscillons} 
%%%%%%%%%%%%%%%%%%%%%%%%%%%%%

At the classical level, oscillons are regarded as stable objects~\cite{Graham:2006xs,Gleiser:2008ty},
although they can decay at the quantum level~\cite{Hertzberg:2010yz}.
%In the present situation, the problem is not so simple because the oscillon solution ceases to exist
%as the cosmic temperature decreases and the thermal potential becomes subdominant.
%Thus oscillons are not absolutely stable in our setup.
Once oscillons are formed, the final decay temperature of $\phi$ condensation may not be estimated
by using the perturbative decay rate of $\phi$.

%There are a couple of effects that makes oscillons unstable.
%(a) Perturbative decay of $\phi$.
%(b) Deformation of the oscillon due to the Hubble expansion.

%%%%%%%%%%%%%%%%%%%%%%%%%%%%%
%\subsubsection{Perturbative decay} 
%%%%%%%%%%%%%%%%%%%%%%%%%%%%%

Inside oscillons, $\phi$ has a large amplitude of $\phi_i$ and $\chi$ obtains a large mass.
Denoting the scalar field configuration of an oscillon by $\phi(r)$ where $r$ is the radius 
measured from the center of oscillon, the decay rate depends on $r$.
%For $r\lesssim r_c$ where $r_c$ is defined as $\phi(r_c) = \phi_c=T/\lambda$,
%the $\chi$ field decouples from thermal bath.
%For $r\gtrsim r_c$, $\chi$ is so light it can take part in thermal bath.
As shown in Sec.~\ref{sec:dyn}, the decay rate is roughly expressed as 
\begin{equation}
	\Gamma_\phi(r) \sim 
	\begin{cases}
		\displaystyle \frac{b\alpha^2 T^3}{\tilde\phi\phi_{\rm th}} & {\rm for~~}r<r_c \\
		\lambda^2 \alpha T & {\rm for~~}r>r_c.
	\end{cases}
\end{equation}
if $m_\phi \ll \alpha T$ where $r_c$ is defined as $\phi(r_c) = \phi_c=T/\lambda$.
On the other hand, we have
\begin{equation}
	\Gamma_\phi(r) \sim 
	\begin{cases}
		\displaystyle \frac{\alpha^2 m_\phi^3}{\tilde\phi\phi_{\rm th}} & {\rm for~~}r<r_c \\
		\lambda^2 m_\phi/ (8\pi) & {\rm for~~}r>r_c.
	\end{cases}
\end{equation}
if $m_\phi \gg gT$ where $r_c$ is defined as $\phi(r_c) = \phi_c=m_\phi/\lambda$.
There may be efficient non-perturbative particle production inside the oscillon as noted in Sec.~\ref{sec:dyn},
but this does not change the results much as long as the parametric resonance does not occur.
This shows that the decay of oscillon occurs most efficiently from the surface at $r\sim r_c$.
Hereafter, we make use of the approximation of $r_c\sim R$, since the typical length scale of the change of the field value is $R$.
The oscillon charge evaporation rate is then roughly approximated by
\begin{equation}
	- \frac{dI}{dt} = \int dr 4\pi r^2  m_\phi \phi(r)^2 \Gamma_\phi(r) \sim \frac{4\pi R^3}{3}m_\phi \phi_i^2 \Gamma_\phi (R).
\end{equation}
The effective oscillon decay rate is given by
\begin{equation}
	\Gamma_I = - \frac{1}{I}\frac{dI}{dt} \sim \frac{1}{\beta}\Gamma_\phi (R)
	\sim 
	\begin{cases}
	\beta^{-1} b\alpha^2 T^3/(\phi_i \phi_{\rm th}) & {\rm for~~}m_\phi \ll \alpha T  \\
	\beta^{-1} \alpha^2 m_\phi^3/(\phi_i \phi_{\rm th}) & {\rm for~~} m_\phi \gg gT.
	\end{cases}
\end{equation}
Note that $\phi_i$ in this expression is the field amplitude inside the oscillon, which does not decrease due to the Hubble friction 
and hence the evaporation epoch is delayed compared with the case without oscillon formation.

%Thus this implies that the $\Gamma_I$ is smaller than $H$ unless $\beta$ is so small
%or $(\tilde\phi/\phi_c)$ is so large.
%As the temperature decreases, the zero-temperature decay rate becomes dominant and then it may exceed the Hubble rate.
%Before then, the oscillon may be deformed as explained below.

%%%%%%%%%%%%%%%%%%%%%%%%%%%%%
%\subsubsection{Deformation} 
%%%%%%%%%%%%%%%%%%%%%%%%%%%%%

%Since the cosmic temperature decreases as the Universe expands, the scalar potential 
%will finally be dominated by the zero-temperature one.
%Then a stable oscillon solution does not exist and oscillons are expected to disappear.
%This happens when the temperature drops down to $T\sim T_{\rm def}= (m_\phi \phi_i/\alpha)^{1/2}$.
%In terms of Hubble parameter, it corresponds to $H\sim H_{\rm def}$ given by
%%
%\begin{equation}
%	H_{\rm def} = \begin{cases}
%	\displaystyle
%	\frac{m_\phi^2\phi_i^2}{\alpha^2 T_{\rm R}^2 M_P} \sim \frac{m_\phi^2}{H_{\rm os}} & {\rm for~~} T_{\rm def} > T_{\rm R} \\
%	\displaystyle
%	\frac{m_\phi \phi_i}{\alpha M_P} & {\rm for~~} T_{\rm def} < T_{\rm R}
%	\end{cases}.
%\end{equation}
%%
%After the oscillon deformation, $\phi$ dissipates its energy into thermal bath in the way described already.
%If the oscillon survives long enough, it is possible that oscillons dominate the Universe
%since the oscillons behave as matter.

%%%%%%%%%%%%%%%%%%%%%%%%%%%%%
\section{Evolution of oscillating scalar fields} 
\label{sec:ev_osc_scalar}
%%%%%%%%%%%%%%%%%%%%%%%%%%%%%

In order to follow the evolution of oscillating scalar fields, 
it is convenient to consider averaged quantities with a time interval which
is longer than the oscillation period but shorter than the Hubble time scale.
In particular, we will derive the formulas which describe the
evolution of $\phi$'s amplitude and 
the evolution of $\phi$'s energy density, and 
clarify the effects of dissipation on them.

%%%%%%%%%%%%%%%%%%%%%%%%%%%%%%%%%%%%%%%
\subsection{Relevant equations}
%%%%%%%%%%%%%%%%%%%%%%%%%%%%%%%%%%%%%%%

Let us consider scalar fields oscillating in the %effective 
potential $V(\phi; T) = m_\phi^2 \phi^2 + V_{\rm eff} (\phi; T)$, 
which includes thermal effects. In the presence of dissipation,
The equation of motion is given by
\begin{align}
	\ddot \phi + ( 3H + \Gamma_\phi )\dot\phi  + \frac{\der V}{\der \phi} = 0.
\end{align}
%%
%where $V' \equiv \partial V/\partial \phi$.
From this, we immediately find the following relation 
between the kinetic energy $K=\dot\phi^2/2$ and the potential energy $V$:
\begin{equation}
	\frac{d}{dt}(K+V) = - ( 6H + 2\Gamma_\phi )K +
	 \frac{\dot T}{T}\left(T\frac{\partial V}{\partial T}\right).   
\end{equation}
This is an exact relation. This means that $K+V$ is changing slowly $O(H,\Gamma_\phi)$,
and approximately conserves with time.
Therefore, we can take time average of this equation as
\begin{align}
	\frac{d}{dt}\langle \overline{K+V}\rangle 
	= -6H\langle \overline K \rangle - 2 \< \overline{ \Gamma_\phi K } \>
	+ \frac{\dot T}{T}\left< \overline {T\frac{\partial V}{\partial T}} \right>, \label{ddtKV}
\end{align}
with a time interval which is longer than the oscillation period but
shorter than the Hubble time scale.
In this section, the time average is denoted as $\< \overline \cdots \>$.

First, we make use of the virial theorem for deriving the averaged motion of $\phi$.
By noting that $2K = d/dt(\phi\dot\phi)-\phi\ddot\phi$, %we take time average of this quantity:
%%
%\begin{equation}
%	\langle 2K\rangle \equiv \lim_{t\to\infty}\frac{1}{t}\int_0^t dt' 2K 
%	= \lim_{t\to\infty}\frac{1}{t}\left[\phi\dot\phi \right]_0^t - \langle \phi\ddot\phi\rangle.
%\e\nd{equation}
%%
%Since the $\phi$ motion is periodic, the first term vanishes in the large $t$ limit. 
%We also note that the oscillation period is much shorter than the Hubble time scale.
%Thus 
we have
\begin{equation}
	\langle \overline{2K}\rangle = \left< \overline{ \phi \frac{\der V}{\der \phi}} \right>,  \label{virial}
\end{equation}
with neglecting $O(H^2,\Gamma_\phi H)$ terms.
This is the virial theorem for an oscillating scalar field.

Next, let us evaluate $\< \overline{2 \Gamma_\phi K} \>$ with some examples. \\

(i) Zero temperature mass at the large field value regime: 
In this case, the $\phi$ dependence of dissipation coefficient is given by $1/ \phi^2$
above the threshold value $\phi_{\rm th}$ which should be determined case by case.
Then one finds
\begin{align}
	\< \overline{\Gamma_\phi \dot\phi^2} \> 
	\propto & \ \left< \overline{ \frac{m_\phi^2}{\tan^2 m_\phi t} } \right> 
	\simeq
	m_\phi^3 \int_{t_{\rm th}} d\tau\ \frac{1}{\tan^2 m_\phi \tau} 
	\sim
	m_\phi \frac{1}{t_{\rm th}} \nn
	\sim &\ 
	\frac{1}{\phi_{\rm th}\tilde\phi} \< \overline{\dot{\phi}^2} \>
\end{align}
Here we have used $\phi \propto \tilde\phi \sin m_\phi t$,
$\phi_{\rm th} \sim \tilde\phi m_\phi t_{\rm th}$
and the virial theorem $\<\overline K\> = \<\overline V\>$.\\

(ii) Thermal logarithmic potential: In this case, the dissipative coefficient is given by
\begin{align}
	\Gamma_\phi \simeq b\,\alpha^2 \frac{T^3 }{\phi^2}.
\end{align}
Hence we  obtain
\begin{align}
	\< \overline{\Gamma_\phi \dot\phi^2} \> 
	= b\, \alpha^2 T^3 \left< \overline{ \frac{\dot\phi^2}{\phi^2}} \right>
	= b\, \alpha^2 T^3 \left< \overline{\frac{\ln (\phi_i^2 / \phi^2)}{\phi^2}} \right> 
	\< \overline{ \dot\phi^2 } \>.
\end{align}
Here we have used the energy conservation 
$\dot\phi^2 / 2 = \alpha^2 T^4 \ln (\phi_i^2 / \phi^2)$ and the virial theorem
$\< \overline{ \dot\phi^2 } \> = 2 \alpha^2 T^4$.
Therefore, the averaged dissipative coefficient is obtained 
\begin{align}
	b\,\alpha^2 T^3 \left< \overline{ \frac{\ln (\phi_i^2 / \phi^2)}{\phi^2} } \right>.
\end{align}
\\

In the following, we denote the averaged dissipative coefficient as $\Gamma_\phi^{\rm eff}$
collectively. Thus, Eq.\ \eqref{ddtKV} can be expressed as
\begin{align}
	\frac{d}{dt} \< \overline{K + V} \>
	=
	- ( 6 H + 2  \Gamma_\phi^{\rm eff} ) \<\overline K \> 
	+ \frac{\dot{T}}{T} \left< \overline{T\frac{\der V}{\der T}} \right>.
	\label{ddtKV2}
\end{align}
From Eqs.~(\ref{ddtKV2}) and (\ref{virial}), 
we obtain a time-averaged relation for the scalar potential as
\begin{align}
	\frac{d}{dt} \left< \overline{\phi \frac{\der V}{\der \phi} +2V} \right> 
	= - ( 6 H + 2  \Gamma_\phi^{\rm eff} ) \left< \overline{\phi \frac{ \der V}{\der \phi}} \right> 
	+ \frac{2\dot T}{T}\left< \overline{T\frac{\partial V}{\partial T}} \right>.
	\label{dVdt}
\end{align}
%%

%%%%%%%%%%%%%%%%%%%%%%%%%%%%%%%%%%%%%%%
\subsection{Evolution of scalar amplitudes in general effective potential}
%%%%%%%%%%%%%%%%%%%%%%%%%%%%%%%%%%%%%%%

Let us derive the evolution of scalar amplitude with
some examples using Eq.\ \eqref{dVdt}.\\

(i) Zero temperature mass: $V=(1/2)m_\phi^2\phi^2$. 
Substituting it into (\ref{dVdt}), we obtain
\begin{equation}
	\frac{d}{dt} \langle \overline {\phi^2}\rangle 
	= - (3H + \Gamma_\phi^{\rm eff}) \langle \overline {\phi^2}\rangle.
	\label{amp_zero}
\end{equation}
Writing $\langle \overline {\phi^2}\rangle = \tilde\phi^2$, we have $\tilde \phi^2\propto a^{-3}$
at the regime $\Gamma_\phi^{\rm eff} \ll H$.\footnote{
	The definition of $\tilde\phi$ here is a bit different from that used in Sec.~\ref{sec:non}.
	The difference, however, is at most $\mathcal O(1)$ and we do not distinguish them
	since our discussions do not require $\mathcal O(1)$ accuracy.
}
\\

(ii) Thermal mass: $V=(1/2)\lambda^2 T^2\phi^2$.
Substituting it into (\ref{dVdt}), we obtain
\begin{equation}
	\frac{d}{dt} \langle \overline {\phi^2}\rangle 
	= -\left(3H+\Gamma_\phi^{\rm eff} +\frac{\dot T}{T}\right)  \langle \overline {\phi^2}\rangle.
	\label{amp_th_mass}
\end{equation}
Thus we have $\tilde\phi^2 \propto a^{-3}T^{-1}$
at the regime $\Gamma_\phi^{\rm eff} \ll H$.\\

(iii) Thermal log: $V=\alpha^2 T^4 \ln (\lambda^2\phi^2/T^2)$.
Substituting it into (\ref{dVdt}), we obtain
\begin{equation}
	\frac{d}{dt}\langle \overline{T^4(1+ \ln\phi^2/T^2)} \rangle 
	= -(6H + 2 \Gamma_\phi^{\rm eff} )T^4
	+\frac{4\dot T}{T}\langle \overline{T^4\ln\phi^2/T^2} \rangle
	- \frac{2 \dot T}{T} T^4.
\end{equation}
The l.h.s. can be written as
\begin{equation}
	\frac{4\dot T}{T}\langle T^4(1+ \ln\phi^2) \rangle + T^4 \frac{d{\tilde\phi^2}/dt}{ \tilde\phi^2 }
	- \frac{2 \dot T}{T} T^4
\end{equation}
where we have defined $\langle \overline{ \ln \phi^2} \rangle \equiv \ln \tilde\phi^2$.
Thus we have
\begin{equation}
	\frac{d}{dt}  \tilde\phi^2 = 
	- \left(6H + 2 \Gamma_\phi^{\rm eff} + \frac{4\dot T}{T}\right)\tilde\phi^2.
	\label{amp_th_log}
\end{equation}
Thus we have $\tilde\phi^2 \propto a^{-6}T^{-4}$ at the regime 
$\Gamma_\phi^{\rm eff} \ll H$.

%%%%%%%%%%%%%%%%%%%%%%%%%%%%%%%%%%%%%%%
\subsection{Evolution of the energy density of scalar field}
%%%%%%%%%%%%%%%%%%%%%%%%%%%%%%%%%%%%%%%

Let us derive the evolution of $\phi$'s energy density, which is defined as
\begin{align}
	\rho_\phi = 
	\left< \overline{\frac{1}{2} \dot\phi^2 + \frac{1}{2} m_\phi^2 \phi^2} \right> 
	= \<\overline K \> + \frac{1}{2} m_\phi^2 \< \overline {\phi^2} \>.
\end{align}
To make our discussion concrete, we will study the evolution of energy density 
with some examples in the following.
Note that the evolution of the latter term, i.e. the potential energy, 
have been obtained in the last section.\\

(i) Zero temperature mass: $V=(1/2)m_\phi^2\phi^2$. 
Using the virial theorem: Eq.\ \eqref{virial},
we obtain
\begin{align}
	\< \overline K \> = \frac{1}{2} m_\phi^2 \< \overline {\phi^2} \>.
\end{align}
Thus, the evolution of $\< \overline K \>$ is the same as 
one of $\< \overline {\phi^2}\>$ given by Eq.\ \eqref{amp_zero}:
\begin{align}
	\frac{d}{dt} \rho_\phi =
	\frac{d}{dt} \< \overline{2K} \> = m_\phi^2 \frac{d}{dt}  \< \overline {\phi^2}\> 
	= - ( 3H + \Gamma_\phi^{\rm eff} ) \rho_\phi.
\end{align}
\\

(ii) Thermal mass: $V=(1/2)\lambda^2 T^2\phi^2$.
The virial theorem implies
\begin{align}
	\< \overline K \> = \frac{1}{2} \lambda^2 T ^2 \< \overline {\phi^2} \>.
\end{align}
As can be seen from this equation, if the $\phi$ oscillates with the thermal mass potential,
then $\< \overline K \> \gg m_\phi^2 \< \overline {\phi^2} \> / 2$ is satisfied.
Differentiating this with respect to time,
we find
\begin{align}
	\frac{d}{dt} \< \overline K \>
	= - \( 3 H + \Gamma_\phi^{\rm eff} - \frac{\dot T}{T} \) \< \overline K \>.
\end{align}
And we also have Eq.\ \eqref{amp_th_mass}:
\begin{align}
	\frac{d}{dt} \left< \overline{\frac{1}{2} m_\phi^2 \phi^2} \right>
	= - \( 3 H +\Gamma_\phi^{\rm eff} + \frac{\dot T}{T} \) 
	\left< \overline{ \frac{1}{2} m_\phi^2 \phi^2 }\right>.
\end{align}
\\

(iii) Thermal log: $V=\alpha^2 T^4 \ln ( \lambda^2\phi^2/T^2 )$.
The virial theorem implies that the kinetic energy is independent of 
the dissipative coefficient:
\begin{align}
	\< \overline K \> = \alpha^2 T^4.
\end{align}
Hence we have
\begin{align}
	\frac{d}{dt} \< \overline K \> = 4 \frac{\dot T}{T} \< \overline K \>,
\end{align}
and Eq.\ \eqref{amp_th_log} means
\begin{align}
	\frac{d}{dt} \left< \overline {\frac{1}{2} m_\phi^2 \phi^2} \right>
	= - \(6 H + 2 \Gamma_\phi^{\rm eff} + \frac{4 \dot T}{T} \) 
	\left< \overline {\frac{1}{2} m_\phi^2 \phi^2} \right>.
\end{align}
Note that the kinetic energy is larger than the potential energy if
the $\phi$ oscillates with the thermal logarithmic potential:
$
	\< \overline K \> \gg \< \overline {m_\phi^2 \phi^2 / 2} \>.
$

%%%%%%%%%%%%%%%%%%%%%%%%%%%%%
\section{Closed Time Path formalism}
\label{sec:CTP} 
%%%%%%%%%%%%%%%%%%%%%%%%%%%%%

In this section, let us briefly introduce the basic ingredients of Closed Time Path
formalism. Though equations and formulas shown in the following
can be found in~\cite{Berges:2004yj,CalzettaHuBook,BellacBook,KapustaBook}
and references therein, we will summarize them for the sake of readers.

%%%%%%%%%%%%%%%%%%%%%%%%%%%%%
\subsection{Closed Time Path formalism}
\label{sec:} 
%%%%%%%%%%%%%%%%%%%%%%%%%%%%%

To follow quantum dynamics, 
we are often interested in the time evolution of expectation value of operators
for a system described by a density matrix $\hat\rho$ at an initial time.
Such expectation values can be calculated in terms of so-called
Closed Time Path (CTP) or in-in formalism (See \cite{Berges:2004yj,CalzettaHuBook} and 
references therein).
The evolution of expectation value of a Heisenberg operator 
can be written in terms of CTP as
\begin{align}
	\< \hat O_{H} (t) \> 
	= \tr \left[ \hat\rho \, T_C \exp \( - i \int_C dt' H_I (t') \) \hat O_I (t) \right]
\end{align}
where the $H$ and $I$ subscript denote the Heisenberg and Interaction 
picture respectively,
the time integration is performed on the CTP contour $C$,
and $T_C$ denotes the contour $C$ ordering.
The Schwinger-Keldysh propagator is defined as
a connected two point correlator with the CTP contour:
\begin{align}
	G (x,y) :=& \ \< T_C \, \hat\varphi (x) \hat\varphi (y) \> - \phi (x) \phi (y) \\
	S(x,y) := & \ \< T_C \, \hat\psi (x) \hat{\bar\psi} (y) \>,
\end{align}
where $\phi(x) := \< \hat\varphi (x) \>$. Two convenient combinations
of propagator are introduced: the Jordan propagator (or the spectral function)
and the Hadamard propagator (or the statistical function).
They are given as follows respectively.
\begin{align}
	G_J (x,y) := & \  \< [ \hat\varphi(x), \hat\varphi (y) ] \>; \ \ 
	G_H (x,y) :=    \< \{ \hat\varphi (x), \hat\varphi (y) \} \> -2\phi(x) \phi(y)  \\
	S_J (x,y) := & \ \< \{ \hat\psi (x), \hat{\bar\psi} (y) \} \>; \ \
	S_H (x,y) :=  \< [ \hat\psi (x), \hat{\bar\psi} (y) ] \>.
\end{align}
Using these propagators, one can express the Schwinger-Keldysh propagator 
as
\begin{align}
	G (x,y) = & \ \frac{1}{2} \left[ G_H (x,y) 
	+ {\rm sgn}_C (x_0,y_0) \, G_J (x,y) \right], \\
	S (x,y) = & \ \frac{1}{2} \left[ S_H (x,y)
	+ {\rm sgn}_C  (x_0,y_0) \, S_J (x,y)
	\right]
\end{align}
where the sign function ${\rm sgn}_C$ is defined on the contour $C$.
It is often convenient to define the retarded and advanced propagators:
\begin{align}
	G_{\rm ret/adv} (x,y) := & \ \pm i \theta (\pm x_0 \mp y_0)\, G_J (x,y) \label{eq:prop_ret_bos} \\
	S_{\rm ret/adv} (x,y) := & \ \pm i \theta (\pm x_0 \mp y_0)\, S_J (x,y). \label{eq:prop_ret_fer}
\end{align}

Note that if a system has a spacial translational invariance,
all the above two point correlators only depend on the difference
of spacial coordinate, ${\bf x} - {\bf y}$.

%%%%%%%%%%%%%%%%%%%%%%%%%%%%%
\subsection{Thermal equilibrium}
\label{sec:thermal_eq} 
%%%%%%%%%%%%%%%%%%%%%%%%%%%%%

In general, the Hadamard and Jordan propagators are independent.
However, if some fields are in thermal equilibrium 
(See \cite{BellacBook,KapustaBook} and references therein), 
their Hadamard and Jordan propagators are related through the Kubo-Martin-Schwinger (KMS)
relation \cite{Kubo:1957mj+x}.
Due to the translational invariance of thermal equilibrium system,
all the two point correlators in thermal equilibrium depend
only relative coordinates and it is convenient to consider the 
Fourier transformations of them:
\begin{align}
	G^{\rm th}_\bullet (x-y) = \int \frac{d^4 k}{( 2\pi )^4}\
	e^{- i k \cdot (x - y) } \ G_\bullet^{\rm th} (k).
\end{align}
The KMS relation implies
\begin{align}
	G_H^{\rm th} (\omega,{\bf k}) 
	=& \ \left[ 1 + 2 f_B (\omega) \right] \rho^B (\omega,{\bf k}), \\
	S_H^{\rm th} (\omega,{\bf k})
	=& \ \left[ 1 - 2 f_F (\omega) \right] \rho^F (\omega,{\bf k}), 
\end{align}
where $f_{B/F}$ is the Bose-Einstein/Fermi-Dirac distribution and
$\rho^{B/F}$ denotes the spectral densities, defined as 
$\rho^{B/F} (\omega, {\bf p}) := G_J^{\rm th}/S_J^{\rm th} (\omega,{\bf p})$ respectively.

For simplicity, the ``${\rm th}$'' superscript is suppressed in the following.
Eqs.\ \eqref{eq:prop_ret_bos} and \eqref{eq:prop_ret_fer} imply 
the following relations:
\begin{align}
	\left.
	\begin{cases}
	G_{\rm ret/adv} (p_0,{\bf p}) \\ S_{\rm ret/adv} (p_0,{\bf p}) 
	\end{cases}
	\!\!\!\!\!\!\!\right\}
	= {\rm PV} \int \frac{dk_0}{2\pi} \frac{1}{p_0 - k_0}
	\left.
	\begin{cases}
	 G_J (k_0,{\bf p}) \\ 
	 S_J (k_0,{\bf p})
	 \end{cases}
	\!\!\!\!\!\!\!\right\}
	\pm \frac{i}{2}
	\left.
	\begin{cases}
	 G_J (p_0,{\bf p}) \\ 
	 S_J (p_0,{\bf p})
	 \end{cases}
	\!\!\!\!\!\!\!\right\}.
\end{align}
The retarded and advanced self energies are defined as
\begin{align}
	\left.
	\begin{cases}
	\Pi_{\rm ret/adv} (x) \\ \Sigma_{\rm ret/adv} (x) 
	\end{cases}
	\!\!\!\!\!\!\!\right\}
	:= \mp i \theta (\pm x_0)\,
	\left.
	\begin{cases}
	 \Pi_J (x) \\ 
	 \Sigma_J (x)
	 \end{cases}
	\!\!\!\!\!\!\!\right\}
\end{align}
and these imply the following relations:
\begin{align}
	\left.
	\begin{cases}
	\Pi_{\rm ret/adv} (p_0,{\bf p}) \\ \Sigma_{\rm ret/adv} (p_0,{\bf p}) 
	\end{cases}
	\!\!\!\!\!\!\!\right\}
	= {\rm PV} \int \frac{dk_0}{2\pi} \frac{1}{p_0 - k_0}
	\left.
	\begin{cases}
	 \Pi_J (k_0,{\bf p}) \\ 
	 \Sigma_J (k_0,{\bf p})
	 \end{cases}
	\!\!\!\!\!\!\!\right\}
	\mp \frac{i}{2}
	\left.
	\begin{cases}
	 \Pi_J (p_0,{\bf p}) \\ 
	 \Sigma_J (p_0,{\bf p})
	 \end{cases}
	\!\!\!\!\!\!\!\right\}.\label{eq:KK_self}
\end{align}

If a real scalar field $\varphi$ interacts with thermal bath by 
${\cal L}_{\rm int} = \varphi \hat O$, then the self energy of $\varphi$
is given by
\begin{align}
	\Pi_J (x) = \< [ \hat O (x), \hat O (0) ] \>
\end{align}
at the leading order in ${\cal L}_{\rm int}$. Here we have assumed that
the back reaction on the thermal bath is negligible.
In this case, the real and imaginary parts of retarded self energy can be 
expressed as
\begin{align}
	\Re \Pi_{\rm ret} (p_0 ,{\bf p}) = &\ 
	{\rm PV} \int \frac{dk_0}{2\pi} \frac{\Pi_J (k_0,{\bf p})}{p_0 - k_0} \\
	\Im \Pi_{\rm ret} (p_0,{\bf p}) = & \
	 - \frac{\Pi_J (p_0,{\bf p})}{2}.
\end{align}
%%

%%%%%%%%%%%%%%%%%%%%%%%%%%%%%
%\subsection{Spectral density}
%\label{sec:} 
%%%%%%%%%%%%%%%%%%%%%%%%%%%%%

In an interacting theory, these spectral densities can be expressed as
\begin{align}
	\rho^B (p_0,{\bf p}) 
	= & \ \frac{1}{i} 
	\left[
		\frac{1}{ m^2 - p^2 +\Pi_{\rm ret} (p_0,{\bf p}) }
		- \frac{1}{ m^2 - p^2 +\Pi_{\rm adv} (p_0,{\bf p}) }  
	\right], \\
	\rho^F (p_0,{\bf p})
	= & \ \frac{1}{i}
	\left[
		\frac{1}{ m_D - \Slash{p}  + \Sigma_{\rm ret} (p_0,{\bf p})}
		- \frac{1}{ m_D - \Slash{p}  + \Sigma_{\rm adv} (p_0,{\bf p})}
	\right],
\end{align}
where $\Pi$ and $\Sigma$ are the self energies for boson and fermion,
and $m_D$ denotes the Dirac mass (See \cite{Anisimov:2010dk} for Majorana fermion).

%%%%%%%%%%%%%%%%%%%%%%%%%%%%%
\subsubsection{Breit-Wigner approximation}
\label{sec:} 
%%%%%%%%%%%%%%%%%%%%%%%%%%%%%

If the spectral densities well concentrate around the poles,
we may approximate them by the Breit-Wigner form.
For a real scalar field, the Breit-Wigner form is given by
\begin{align}
	\rho^B (p_0,{\bf p})
	= \frac{2 p_0 \Gamma_{\bf p}}{[p_0^2 - \Omega_{\bf p}^2]^2 + [p_0 \Gamma_{\bf p}]^2},
\end{align}
where $\Omega_{\bf p} = \sqrt {m^2 + {\bf p}^2 + \Re\Pi_{\rm ret} (\Omega_{\bf p},{\bf p})}$
and $\Gamma_{\bf p} = - \Im\Pi_{\rm ret} (\Omega_{\bf p},{\bf p}) / \Omega_{\bf p}$.

As shown in Ref.\ \cite{Wang:1999mb}, the Breit-Wigner form of spectral density 
for fermion with the vanishing mass term
can be expressed as
\begin{align}
	&\left. \rho^F (p_0,{\bf p})\right|_{m_D \simeq 0} \nn
	& = \sum_{s = \pm} \frac{Z_{\bf p}^s}{2} \left[
		\frac{\Gamma_{\bf p}^s}{[p_0 - \Omega^s_{\bf p}]^2 + \Gamma_{\bf p}^s{^2}/4 }
		 \( \gamma_0 - \hat{\bf q} \cdot  \mbox{\boldmath{$\gamma$}} \) +
		\frac{\Gamma_{\bf p}^s}{[p_0 + \Omega^s_{\bf p}]^2 + \Gamma_{\bf p}^s{^2}/4 }
		 \( \gamma_0 + \hat{\bf q} \cdot  \mbox{\boldmath{$\gamma$}} \)
	\right]
\end{align}
where the real and imaginary parts of the pole are determined by
\begin{align}
	\Omega^\pm_{\bf p} = & \ \pm p - A^R_{\bf p} (\Omega_{\bf p}^\pm) 
	\( \Omega^\pm_{\bf p} \mp p \) - B^R_{\bf p} (\Omega_{\bf p}^\pm) \\
	\Gamma^\pm_{\bf p}/2 = & \ 
	Z_{\bf p}^\pm \left[ A^I_{\bf p} (\Omega_{\bf p}^\pm) \( \Omega^\pm_{\bf p} \mp p \)
	+B^I_{\bf p} (\Omega_{\bf p}^\pm) \right]
\end{align}
and the wave functional renormalization is given by
\
\begin{align}
	Z^\pm_{\bf p} = \left(1+
	\left. \frac{\der }{\der \omega} \left[  
	A^R_{\bf p}(\omega) (\omega \mp p) + B^R_{\bf p} (\omega)
	\right] \right|_{\omega = \Omega_{\bf p}^\pm} \right)^{-1}.
\end{align}
From Eq.\ \eqref{eq:KK_self}, the self energies can be decomposed into 
$\Sigma_{\rm ret/adv}=\Sigma \mp i \Sigma_J / 2$, and can be
expressed as
\begin{align}
	\Sigma (p_0,{\bf p}) 
	= & \ - A^R_{\bf p} (p_0)\, \Slash{p} - B^R_{\bf p} (p_0)\, \gamma_0 \\
	\Sigma_J (p_0,{\bf p})/2 
	= & \ A^I_{\bf p} (p_0)\, \Slash{p} + B^I_{\bf p} (p_0)\, \gamma_0.
\end{align}

On the other hand, as shown in Ref.\ \cite{BasteroGil:2010pb},
if the zero temperature mass term is much larger than 
the typical thermal mass scale $gT$,
then the Breit-Wigner form of spectral density for fermion can be approximated by
\begin{align}
	\left. \rho^F (p_0,{\bf p}) \right|_{m_D \gg gT}
	= (m_D + \Slash{p})
	\frac{2p_0 \Gamma_{\bf p}}{[p_0^2 - \omega_{\bf p}^2]^2 + [p_0\Gamma_{\bf p}]^2}
\end{align}
where the real and imaginary parts of the pole are determined by
\begin{align}
	\omega_{\bf p} = & \ \sqrt{m_D^2 + {\bf p}^2}\\
	\Gamma_{\bf p}/2 = & \ 
	\frac{1}{\omega_{\bf p}}
	\left[
		m_D^2 \( A^I_{\bf p} (\omega_{\bf p}) + C^I_{\bf p}(\omega_{\bf p}) \)
		+B_{\bf p}^I (\omega_{\bf p}) \omega_{\bf p}
	\right].
\end{align}
In the presence of Dirac mass term which breaks the chiral symmetry,
the self energies can be expressed as
\begin{align}
	\Sigma (p_0,{\bf p}) 
	= & \ - A^R_{\bf p} (p_0)\, \Slash{p} - B^R_{\bf p} (p_0)\, \gamma_0 - 
	C^R_{\bf p} (p_0) \, m_D\\
	\Sigma_J (p_0,{\bf p})/2 
	= & \ A^I_{\bf p} (p_0)\, \Slash{p} + B^I_{\bf p} (p_0)\, \gamma_0
	+ C^I_{\bf p} (p_0) \, m_D.
\end{align}
%%

%%%%%%%%%%%%%%%%%%%%%%%%%%%%%%%%%%%%%%%%%%%%

%%%%%%%%%%%%%%%%%%%%%%%%%%%%%%%%%%%%%%%%%%%%


\begin{thebibliography}{99}
%%%%%%%%%%%%%%%%%%%%%%%%%%%%%%%%%%%%%%%%%%%%



   %\cite{Affleck:1984fy}
\bibitem{Affleck:1984fy} 
  I.~Affleck and M.~Dine,
  %``A New Mechanism for Baryogenesis,''
  Nucl.\ Phys.\ B {\bf 249}, 361 (1985).
  %%CITATION = NUPHA,B249,361;%%

%\cite{Lyth:2009zz}
\bibitem{Lyth:2009zz} 
  See e.g., D.~H.~Lyth and A.~R.~Liddle,
  ``The primordial density perturbation: Cosmology, inflation and the origin of structure,''
  Cambridge, UK: Cambridge Univ. Pr. (2009) 497 p
  
  %\cite{Mollerach:1989hu}
\bibitem{Mollerach:1989hu} 
  S.~Mollerach,
  %``Isocurvature Baryon Perturbations And Inflation,''
  Phys.\ Rev.\ D {\bf 42}, 313 (1990).
  %%CITATION = PHRVA,D42,313;%%
  
  %\cite{Linde:1996gt}
\bibitem{Linde:1996gt} 
  A.~D.~Linde and V.~F.~Mukhanov,
  %``Nongaussian isocurvature perturbations from inflation,''
  Phys.\ Rev.\ D {\bf 56}, 535 (1997)
  [astro-ph/9610219].
  %%CITATION = ASTRO-PH/9610219;%%

%\cite{Lyth:2001nq}
\bibitem{Lyth:2001nq} 
  D.~H.~Lyth and D.~Wands,
  %``Generating the curvature perturbation without an inflaton,''
  Phys.\ Lett.\ B {\bf 524}, 5 (2002)
  [hep-ph/0110002].
  %%CITATION = HEP-PH/0110002;%%
  
%\cite{Moroi:2001ct}
\bibitem{Moroi:2001ct} 
  T.~Moroi and T.~Takahashi,
  %``Effects of cosmological moduli fields on cosmic microwave background,''
  Phys.\ Lett.\ B {\bf 522}, 215 (2001)
  [Erratum-ibid.\ B {\bf 539}, 303 (2002)]
  [hep-ph/0110096].
  %%CITATION = HEP-PH/0110096;%%


  
  %\cite{KapustaBook}
\bibitem{KapustaBook} 
  J.~I.~Kapusta and C.~Gale,
  ``Finite-temperature field theory: Principles and applications,''
  Cambridge University Press, Cambridge, UK (2006).
  
  %\cite{Anisimov:2000wx}
\bibitem{Anisimov:2000wx} 
  A.~Anisimov and M.~Dine,
  %``Some issues in flat direction baryogenesis,''
  Nucl.\ Phys.\ B {\bf 619}, 729 (2001)
  [hep-ph/0008058].
  %%CITATION = HEP-PH/0008058;%%
  
  %\cite{Berera:1995ie}
\bibitem{Berera:1995ie} 
  A.~Berera,
  %``Warm inflation,''
  Phys.\ Rev.\ Lett.\  {\bf 75}, 3218 (1995)
  [astro-ph/9509049].
  %%CITATION = ASTRO-PH/9509049;%%
 
   %\cite{Berera:2008ar}
\bibitem{Berera:2008ar} 
  A.~Berera, I.~G.~Moss and R.~O.~Ramos,
  %``Warm Inflation and its Microphysical Basis,''
  Rept.\ Prog.\ Phys.\  {\bf 72}, 026901 (2009)
  [arXiv:0808.1855 [hep-ph]].
  %%CITATION = ARXIV:0808.1855;%%
  
  %\cite{BasteroGil:2009ec}
\bibitem{BasteroGil:2009ec} 
  M.~Bastero-Gil and A.~Berera,
  %``Warm inflation model building,''
  Int.\ J.\ Mod.\ Phys.\ A {\bf 24}, 2207 (2009)
  [arXiv:0902.0521 [hep-ph]].
  %%CITATION = ARXIV:0902.0521;%%
    
  %\cite{Yokoyama:2004pf}
\bibitem{Yokoyama:2004pf} 
  J.~'i.~Yokoyama,
  %``Fate of oscillating scalar fields in the thermal bath and their cosmological implications,''
  Phys.\ Rev.\ D {\bf 70}, 103511 (2004)
  [hep-ph/0406072].
  %%CITATION = HEP-PH/0406072;%%
  
%\cite{Drewes:2010pf}
\bibitem{Drewes:2010pf} 
  M.~Drewes,
  %``On the Role of Quasiparticles and thermal Masses in Nonequilibrium Processes in a Plasma,''
  arXiv:1012.5380 [hep-th].
  %%CITATION = ARXIV:1012.5380;%%
  
  %\cite{BasteroGil:2010pb}
\bibitem{BasteroGil:2010pb} 
  M.~Bastero-Gil, A.~Berera and R.~O.~Ramos,
  %``Dissipation coefficients from scalar and fermion quantum field interactions,''
  JCAP {\bf 1109}, 033 (2011)
  [arXiv:1008.1929 [hep-ph]].
  %%CITATION = ARXIV:1008.1929;%%


  %\cite{Kofman:1994rk}
\bibitem{Kofman:1994rk} 
  L.~Kofman, A.~D.~Linde and A.~A.~Starobinsky,
  %``Reheating after inflation,''
  Phys.\ Rev.\ Lett.\  {\bf 73}, 3195 (1994)
  [hep-th/9405187].
  %%CITATION = HEP-TH/9405187;%%
  
  %\cite{Shtanov:1994ce}
\bibitem{Shtanov:1994ce} 
  Y.~Shtanov, J.~H.~Traschen and R.~H.~Brandenberger,
  %``Universe reheating after inflation,''
  Phys.\ Rev.\ D {\bf 51}, 5438 (1995)
  [hep-ph/9407247].
  %%CITATION = HEP-PH/9407247;%%
  
  %\cite{Kofman:1997yn}
\bibitem{Kofman:1997yn} 
  L.~Kofman, A.~D.~Linde and A.~A.~Starobinsky,
  %``Towards the theory of reheating after inflation,''
  Phys.\ Rev.\ D {\bf 56}, 3258 (1997)
  [hep-ph/9704452].
  %%CITATION = HEP-PH/9704452;%%
  
     %\cite{Felder:1998vq}
\bibitem{Felder:1998vq} 
  G.~N.~Felder, L.~Kofman and A.~D.~Linde,
  %``Instant preheating,''
  Phys.\ Rev.\ D {\bf 59}, 123523 (1999)
  [hep-ph/9812289].
  %%CITATION = HEP-PH/9812289;%%

  %\cite{Coleman:1973jx}
\bibitem{Coleman:1973jx} 
  S.~R.~Coleman and E.~J.~Weinberg,
  %``Radiative Corrections as the Origin of Spontaneous Symmetry Breaking,''
  Phys.\ Rev.\ D {\bf 7}, 1888 (1973).
  %%CITATION = PHRVA,D7,1888;%%
  
   %\cite{Dine:1995kz}
\bibitem{Dine:1995kz} 
  M.~Dine, L.~Randall and S.~D.~Thomas,
  %``Baryogenesis from flat directions of the supersymmetric standard model,''
  Nucl.\ Phys.\ B {\bf 458}, 291 (1996)
  [hep-ph/9507453].
  %%CITATION = HEP-PH/9507453;%%

  
  %\cite{Lyth:2005fi}
\bibitem{Lyth:2005fi} 
  D.~H.~Lyth and Y.~Rodriguez,
  %``The Inflationary prediction for primordial non-Gaussianity,''
  Phys.\ Rev.\ Lett.\  {\bf 95}, 121302 (2005)
  [astro-ph/0504045].
  %%CITATION = ASTRO-PH/0504045;%%
  
  %\cite{Ichikawa:2008iq}
\bibitem{Ichikawa:2008iq} 
  K.~Ichikawa, T.~Suyama, T.~Takahashi and M.~Yamaguchi,
  %``Non-Gaussianity, Spectral Index and Tensor Modes in Mixed Inflaton and Curvaton Models,''
  Phys.\ Rev.\ D {\bf 78}, 023513 (2008)
  [arXiv:0802.4138 [astro-ph]].
  %%CITATION = ARXIV:0802.4138;%%
  
  %\cite{Suyama:2010uj}
\bibitem{Suyama:2010uj} 
  For a review, see e.g, T.~Suyama, T.~Takahashi, M.~Yamaguchi and S.~Yokoyama,
  %``On Classification of Models of Large Local-Type Non-Gaussianity,''
  JCAP {\bf 1012}, 030 (2010)
  [arXiv:1009.1979 [astro-ph.CO]].
  %%CITATION = ARXIV:1009.1979;%%  
  
  %\cite{Enqvist:2005pg}
\bibitem{Enqvist:2005pg} 
  K.~Enqvist and S.~Nurmi,
  %``Non-gaussianity in curvaton models with nearly quadratic potential,''
  JCAP {\bf 0510}, 013 (2005)
  [astro-ph/0508573];
  %%CITATION = ASTRO-PH/0508573;%%
  %\cite{Enqvist:2008gk}
%\bibitem{Enqvist:2008gk} 
  K.~Enqvist and T.~Takahashi,
  %``Signatures of Non-Gaussianity in the Curvaton Model,''
  JCAP {\bf 0809}, 012 (2008)
  [arXiv:0807.3069 [astro-ph]];
  %%CITATION = ARXIV:0807.3069;%%
  %\cite{Huang:2008zj}
%\bibitem{Huang:2008zj} 
  Q.~-G.~Huang,
  %``Curvaton with Polynomial Potential,''
  JCAP {\bf 0811}, 005 (2008)
  [arXiv:0808.1793 [hep-th]];
  %%CITATION = ARXIV:0808.1793;%%
  %\cite{Enqvist:2009zf}
%\bibitem{Enqvist:2009zf} 
  K.~Enqvist, S.~Nurmi, G.~Rigopoulos, O.~Taanila and T.~Takahashi,
  %``The Subdominant Curvaton,''
  JCAP {\bf 0911}, 003 (2009)
  [arXiv:0906.3126 [astro-ph.CO]];
  %%CITATION = ARXIV:0906.3126;%%
  %\cite{Enqvist:2009ww}
%\bibitem{Enqvist:2009ww} 
  K.~Enqvist, S.~Nurmi, O.~Taanila and T.~Takahashi,
  %``Non-Gaussian Fingerprints of Self-Interacting Curvaton,''
  JCAP {\bf 1004}, 009 (2010)
  [arXiv:0912.4657 [astro-ph.CO]].
  %%CITATION = ARXIV:0912.4657;%%
  
  %\cite{Kawasaki:2008mc}
\bibitem{Kawasaki:2008mc} 
  M.~Kawasaki, K.~Nakayama and F.~Takahashi,
  %``Hilltop Non-Gaussianity,''
  JCAP {\bf 0901}, 026 (2009)
  [arXiv:0810.1585 [hep-ph]];
  %%CITATION = ARXIV:0810.1585;%%
  %\cite{Huang:2010cy}
%\bibitem{Huang:2010cy} 
  Q.~-G.~Huang,
  %``Negative spectral index of $f_{NL}$ in the axion-type curvaton model,''
  JCAP {\bf 1011}, 026 (2010)
  [Erratum-ibid.\  {\bf 1102}, E01 (2011)]
  [arXiv:1008.2641 [astro-ph.CO]];
  %%CITATION = ARXIV:1008.2641;%%
  %\cite{Fonseca:2011aa}
%\bibitem{Fonseca:2011aa} 
  J.~Fonseca and D.~Wands,
  %``Non-Gaussianity and Gravitational Waves from Quadratic and Self-interacting Curvaton,''
  Phys.\ Rev.\ D {\bf 83}, 064025 (2011)
  [arXiv:1101.1254 [astro-ph.CO]];
  %%CITATION = ARXIV:1101.1254;%%
  %\cite{Kawasaki:2011pd}
%\bibitem{Kawasaki:2011pd} 
  M.~Kawasaki, T.~Kobayashi and F.~Takahashi,
  %``Non-Gaussianity from Curvatons Revisited,''
  Phys.\ Rev.\ D {\bf 84}, 123506 (2011)
  [arXiv:1107.6011 [astro-ph.CO]].
  %%CITATION = ARXIV:1107.6011;%%
  
  
  
  %\cite{Gherghetta:1995dv}
\bibitem{Gherghetta:1995dv} 
  T.~Gherghetta, C.~F.~Kolda and S.~P.~Martin,
  %``Flat directions in the scalar potential of the supersymmetric standard model,''
  Nucl.\ Phys.\ B {\bf 468}, 37 (1996)
  [hep-ph/9510370].
  %%CITATION = HEP-PH/9510370;%%
  
  %\cite{Allahverdi:2000zd}
\bibitem{Allahverdi:2000zd} 
  R.~Allahverdi, B.~A.~Campbell and J.~R.~Ellis,
  %``Reheating and supersymmetric flat direction baryogenesis,''
  Nucl.\ Phys.\ B {\bf 579}, 355 (2000)
  [hep-ph/0001122].
  %%CITATION = HEP-PH/0001122;%%

  
  %\cite{Fujii:2001zr}
\bibitem{Fujii:2001zr} 
  M.~Fujii, K.~Hamaguchi and T.~Yanagida,
  %``Reheating temperature independence of cosmological baryon asymmetry in Affleck-Dine leptogenesis,''
  Phys.\ Rev.\ D {\bf 63}, 123513 (2001)
  [hep-ph/0102187].
  %%CITATION = HEP-PH/0102187;%%
  
   %\cite{Coleman:1985ki}
\bibitem{Coleman:1985ki} 
  S.~R.~Coleman,
  %``Q Balls,''
  Nucl.\ Phys.\ B {\bf 262}, 263 (1985)
  [Erratum-ibid.\ B {\bf 269}, 744 (1986)].
  %%CITATION = NUPHA,B262,263;%%
  
  %\cite{Cohen:1986ct}
\bibitem{Cohen:1986ct} 
  A.~G.~Cohen, S.~R.~Coleman, H.~Georgi and A.~Manohar,
  %``The Evaporation Of Q Balls,''
  Nucl.\ Phys.\ B {\bf 272}, 301 (1986).
  %%CITATION = NUPHA,B272,301;%%
  
  %\cite{Kusenko:1997si}
\bibitem{Kusenko:1997si} 
  A.~Kusenko and M.~E.~Shaposhnikov,
  %``Supersymmetric Q balls as dark matter,''
  Phys.\ Lett.\ B {\bf 418}, 46 (1998)
  [hep-ph/9709492].
  %%CITATION = HEP-PH/9709492;%%
  
  %\cite{Enqvist:1997si}
\bibitem{Enqvist:1997si} 
  K.~Enqvist and J.~McDonald,
  %``Q balls and baryogenesis in the MSSM,''
  Phys.\ Lett.\ B {\bf 425}, 309 (1998)
  [hep-ph/9711514];
  %%CITATION = HEP-PH/9711514;%%
  %\cite{Enqvist:1998en}
%\bibitem{Enqvist:1998en} 
  %K.~Enqvist and J.~McDonald,
  %``B - ball baryogenesis and the baryon to dark matter ratio,''
  Nucl.\ Phys.\ B {\bf 538}, 321 (1999)
  [hep-ph/9803380].
  %%CITATION = HEP-PH/9803380;%%
  
  %\cite{Laine:1998rg}
\bibitem{Laine:1998rg} 
  M.~Laine and M.~E.~Shaposhnikov,
  %``Thermodynamics of nontopological solitons,''
  Nucl.\ Phys.\ B {\bf 532}, 376 (1998)
  [hep-ph/9804237].
  %%CITATION = HEP-PH/9804237;%%
  
  %\cite{Banerjee:2000mb}
\bibitem{Banerjee:2000mb} 
  R.~Banerjee and K.~Jedamzik,
  %``On B-ball dark matter and baryogenesis,''
  Phys.\ Lett.\ B {\bf 484}, 278 (2000)
  [hep-ph/0005031].
  %%CITATION = HEP-PH/0005031;%%
  
  %\cite{Kasuya:1999wu}
\bibitem{Kasuya:1999wu} 
  S.~Kasuya and M.~Kawasaki,
  %``Q ball formation through Affleck-Dine mechanism,''
  Phys.\ Rev.\ D {\bf 61}, 041301 (2000)
  [hep-ph/9909509];
  %%CITATION = HEP-PH/9909509;%%
  %\cite{Kasuya:2000sc}
%\bibitem{Kasuya:2000sc} 
  %S.~Kasuya and M.~Kawasaki,
  %``A New type of stable Q balls in the gauge mediated SUSY breaking,''
  Phys.\ Rev.\ Lett.\  {\bf 85}, 2677 (2000)
  [hep-ph/0006128];
  %%CITATION = HEP-PH/0006128;%%
  %\cite{Kasuya:2001hg}
%\bibitem{Kasuya:2001hg} 
  %S.~Kasuya and M.~Kawasaki,
  %``Q ball formation: Obstacle to Affleck-Dine baryogenesis in the gauge mediated SUSY breaking?,''
  Phys.\ Rev.\ D {\bf 64}, 123515 (2001)
  [hep-ph/0106119];
  %%CITATION = HEP-PH/0106119;%%
  %\cite{Hiramatsu:2010dx}
%\bibitem{Hiramatsu:2010dx} 
  T.~Hiramatsu, M.~Kawasaki and F.~Takahashi,
  %``Numerical study of Q-ball formation in gravity mediation,''
  JCAP {\bf 1006}, 008 (2010)
  [arXiv:1003.1779 [hep-ph]].
  %%CITATION = ARXIV:1003.1779;%%
  
  %\cite{Fujii:2001xp}
\bibitem{Fujii:2001xp} 
  M.~Fujii and K.~Hamaguchi,
  %``Higgsino and wino dark matter from Q ball decay,''
  Phys.\ Lett.\ B {\bf 525}, 143 (2002)
  [hep-ph/0110072];
  %%CITATION = HEP-PH/0110072;%%
  %\cite{Fujii:2002kr}
%\bibitem{Fujii:2002kr} 
  %M.~Fujii and K.~Hamaguchi,
  %``Nonthermal dark matter via Affleck-Dine baryogenesis and its detection possibility,''
  Phys.\ Rev.\ D {\bf 66}, 083501 (2002)
  [hep-ph/0205044].
  %%CITATION = HEP-PH/0205044;%%
  
  %\cite{Murayama:1992ua}
\bibitem{Murayama:1992ua} 
  H.~Murayama, H.~Suzuki, T.~Yanagida and J.~'i.~Yokoyama,
  %``Chaotic inflation and baryogenesis by right-handed sneutrinos,''
  Phys.\ Rev.\ Lett.\  {\bf 70}, 1912 (1993);
  %%CITATION = PRLTA,70,1912;%%
  %\cite{Murayama:1993xu}
%\bibitem{Murayama:1993xu} 
  %H.~Murayama, H.~Suzuki, T.~Yanagida and J.~'i.~Yokoyama,
  %``Chaotic inflation and baryogenesis in supergravity,''
  Phys.\ Rev.\ D {\bf 50}, 2356 (1994)
  [hep-ph/9311326].
  %%CITATION = HEP-PH/9311326;%%
  
  %\cite{Murayama:1993em}
\bibitem{Murayama:1993em} 
  H.~Murayama and T.~Yanagida,
  %``Leptogenesis in supersymmetric standard model with right-handed neutrino,''
  Phys.\ Lett.\ B {\bf 322}, 349 (1994)
  [hep-ph/9310297].
  %%CITATION = HEP-PH/9310297;%%
  
  %\cite{Hamaguchi:2001gw}
\bibitem{Hamaguchi:2001gw} 
  K.~Hamaguchi, H.~Murayama and T.~Yanagida,
  %``Leptogenesis from N dominated early universe,''
  Phys.\ Rev.\ D {\bf 65}, 043512 (2002)
  [hep-ph/0109030].
  %%CITATION = HEP-PH/0109030;%%
  
  %\cite{Hamaguchi:2002vc}
\bibitem{Hamaguchi:2002vc} 
  K.~Hamaguchi,
  %``Cosmological baryon asymmetry and neutrinos: Baryogenesis via leptogenesis in supersymmetric theories,''
  hep-ph/0212305.
  %%CITATION = HEP-PH/0212305;%%
  
  %\cite{Moroi:2002vx}
\bibitem{Moroi:2002vx} 
  T.~Moroi and H.~Murayama,
  %``CMB anisotropy from baryogenesis by a scalar field,''
  Phys.\ Lett.\ B {\bf 553}, 126 (2003)
  [hep-ph/0211019].
  %%CITATION = HEP-PH/0211019;%%
  
  %\cite{McDonald:2003xq}
\bibitem{McDonald:2003xq} 
  J.~McDonald,
  %``Right-handed sneutrinos as curvatons,''
  Phys.\ Rev.\ D {\bf 68}, 043505 (2003)
  [hep-ph/0302222];
  %%CITATION = HEP-PH/0302222;%%
  %\cite{McDonald:2004by}
%\bibitem{McDonald:2004by} 
  %J.~McDonald,
  %``Conditions for a successful right-handed Majorana sneutrino curvaton,''
  Phys.\ Rev.\ D {\bf 70}, 063520 (2004)
  [hep-ph/0404154].
  %%CITATION = HEP-PH/0404154;%%
  
  %\cite{Allahverdi:2004ix}
\bibitem{Allahverdi:2004ix} 
  R.~Allahverdi and M.~Drees,
  %``Leptogenesis from a sneutrino condensate revisited,''
  Phys.\ Rev.\ D {\bf 69}, 103522 (2004)
  [hep-ph/0401054].
  %%CITATION = HEP-PH/0401054;%%
  
    %\cite{Peccei:1977hh}
\bibitem{Peccei:1977hh}
  R.~D.~Peccei, H.~R.~Quinn,
  %``CP Conservation in the Presence of Instantons,''
  Phys.\ Rev.\ Lett.\  {\bf 38}, 1440-1443 (1977).
    
  %\cite{Kim:1986ax}
\bibitem{Kim:1986ax}
  For reviews, see J.~E.~Kim,
  %``Light Pseudoscalars, Particle Physics and Cosmology,''
  Phys.\ Rept.\  {\bf 150}, 1 (1987);
  %%CITATION = PRPLC,150,1;%%
  %\cite{Kim:2008hd}
  %\cite{Kim:2008hd}
  %\bibitem{Kim:2008hd}
  J.~E.~Kim and G.~Carosi,
  %``Axions and the Strong CP Problem,''
  Rev.\ Mod.\ Phys.\  {\bf 82}, 557 (2010).
  %[arXiv:0807.3125 [hep-ph]].
  %%CITATION = RMPHA,82,557;%%
  
    %\cite{Kim:1979if}
\bibitem{Kim:1979if}
  J.~E.~Kim,
  %``Weak Interaction Singlet and Strong CP Invariance,''
  Phys.\ Rev.\ Lett.\  {\bf 43}, 103 (1979);
  %\cite{Shifman:1979if}
%\bibitem{Shifman:1979if}
  M.~A.~Shifman, A.~I.~Vainshtein, V.~I.~Zakharov,
  %``Can Confinement Ensure Natural CP Invariance of Strong Interactions?,''
  Nucl.\ Phys.\  {\bf B166}, 493 (1980).

%\cite{Asaka:1998ns}
\bibitem{Asaka:1998ns}
  T.~Asaka and M.~Yamaguchi,
  %``Hadronic axion model in gauge mediated supersymmetry breaking,''
  Phys.\ Lett.\  B {\bf 437}, 51 (1998)
  [arXiv:hep-ph/9805449];
  %%CITATION = PHLTA,B437,51;%%
   %\cite{Asaka:1998xa}
%\bibitem{Asaka:1998xa}
  %T.~Asaka and M.~Yamaguchi,
  %``Hadronic axion model in gauge mediated supersymmetry breaking and cosmology
  %of saxion,''
  Phys.\ Rev.\  D {\bf 59}, 125003 (1999)
  [arXiv:hep-ph/9811451].
  %%CITATION = PHRVA,D59,125003;%%
  
      %\cite{Chun:2000jr}
\bibitem{Chun:2000jr}
  E.~J.~Chun, D.~Comelli and D.~H.~Lyth,
  %``The Abundance of relativistic axions in a flaton model of Peccei-Quinn
  %symmetry,''
  Phys.\ Rev.\  D {\bf 62}, 095013 (2000)
  [arXiv:hep-ph/0008133].
  %%CITATION = PHRVA,D62,095013;%%
  
  %\cite{Abe:2001cg}
\bibitem{Abe:2001cg} 
  N.~Abe, T.~Moroi and M.~Yamaguchi,
  %``Anomaly mediated supersymmetry breaking with axion,''
  JHEP {\bf 0201}, 010 (2002)
  [hep-ph/0111155].
  %%CITATION = HEP-PH/0111155;%%
  
  %\cite{Banks:2002sd}
\bibitem{Banks:2002sd} 
  T.~Banks, M.~Dine and M.~Graesser,
  %``Supersymmetry, axions and cosmology,''
  Phys.\ Rev.\ D {\bf 68}, 075011 (2003)
  [hep-ph/0210256].
  %%CITATION = HEP-PH/0210256;%%
  
  %\cite{Kawasaki:2007mk}
\bibitem{Kawasaki:2007mk} 
  M.~Kawasaki, K.~Nakayama and M.~Senami,
  %``Cosmological implications of supersymmetric axion models,''
  JCAP {\bf 0803}, 009 (2008)
  [arXiv:0711.3083 [hep-ph]].
  %%CITATION = ARXIV:0711.3083;%%
  
  %\cite{Kim:2008yu}
\bibitem{Kim:2008yu}
  S.~Kim, W.~I.~Park and E.~D.~Stewart,
  %``Thermal inflation, baryogenesis and axions,''
  JHEP {\bf 0901}, 015 (2009)
  [arXiv:0807.3607 [hep-ph]].
  %%CITATION = JHEPA,0901,015;%%
  
  %\cite{Choi:2009qd}
\bibitem{Choi:2009qd}
  K.~Choi, K.~S.~Jeong, W.~I.~Park and C.~S.~Shin,
  %``Thermal inflation and baryogenesis in heavy gravitino scenario,''
  JCAP {\bf 0911}, 018 (2009)
  [arXiv:0908.2154 [hep-ph]].
  %%CITATION = JCAPA,0911,018;%%
  
  %\cite{Park:2010qd}
\bibitem{Park:2010qd}
  W.~I.~Park,
  %``A simple model for particle physics and cosmology,''
  JHEP {\bf 1007}, 085 (2010)
  [arXiv:1004.2326 [hep-ph]].
  %%CITATION = JHEPA,1007,085;%%
  
    %\cite{Choi:2011rs}
\bibitem{Choi:2011rs}
  K.~Choi, E.~J.~Chun, H.~D.~Kim, W.~I.~Park and C.~S.~Shin,
  %``The $\mu$-problem and axion in gauge mediation,''
  Phys.\ Rev.\  D {\bf 83}, 123503 (2011)
  [arXiv:1102.2900 [hep-ph]].
  %%CITATION = PHRVA,D83,123503;%%
  
%\cite{Jeong:2011xu}
\bibitem{Jeong:2011xu}
  K.~S.~Jeong and M.~Yamaguchi,
  %``Axion model in gauge-mediated supersymmetry breaking and a solution to the
  %$\mu/B\mu$ problem,''
  JHEP {\bf 1107}, 124 (2011)
  [arXiv:1102.3301 [hep-ph]].
  %%CITATION = JHEPA,1107,124;%%
  
  %\cite{Kawasaki:2010gv}
\bibitem{Kawasaki:2010gv} 
  M.~Kawasaki, N.~Kitajima and K.~Nakayama,
  %``Inflation from a Supersymmetric Axion Model,''
  Phys.\ Rev.\ D {\bf 82}, 123531 (2010)
  [arXiv:1008.5013 [hep-ph]];
  %%CITATION = ARXIV:1008.5013;%%
  %\cite{Kawasaki:2011ym}
%\bibitem{Kawasaki:2011ym} 
  %M.~Kawasaki, N.~Kitajima and K.~Nakayama,
  %``Cosmological Aspects of Inflation in a Supersymmetric Axion Model,''
  Phys.\ Rev.\ D {\bf 83}, 123521 (2011)
  [arXiv:1104.1262 [hep-ph]].
  %%CITATION = ARXIV:1104.1262;%%
  
  %\cite{Kawasaki:2011aa}
\bibitem{Kawasaki:2011aa} 
  M.~Kawasaki, N.~Kitajima and K.~Nakayama,
  %``Revisiting the cosmological coherent oscillation,''
  arXiv:1112.2818 [hep-ph].
  %%CITATION = ARXIV:1112.2818;%%
  
  %\cite{Nakayama:2012zc}
\bibitem{Nakayama:2012zc} 
  K.~Nakayama and N.~Yokozaki,
  %``Peccei-Quinn extended gauge-mediation model with vector-like matter,''
  arXiv:1204.5420 [hep-ph].
  %%CITATION = ARXIV:1204.5420;%%
  
  %\cite{Moroi:2012vu}
\bibitem{Moroi:2012vu} 
  T.~Moroi and M.~Takimoto,
  %``Thermal Effects on Saxion in Supersymmetric Model with Peccei-Quinn Symmetry,''
  arXiv:1207.4858 [hep-ph].
  %%CITATION = ARXIV:1207.4858;%%
  
    %\cite{Yamamoto:1985rd}
\bibitem{Yamamoto:1985rd}
  K.~Yamamoto,
  %``PHASE TRANSITION ASSOCIATED WITH INTERMEDIATE GAUGE SYMMETRY BREAKING IN
  %SUPERSTRING MODELS,''
  Phys.\ Lett.\  B {\bf 168}, 341 (1986);
  %%CITATION = PHLTA,B168,341;%%
  %\cite{Lazarides:1985ja}
%\bibitem{Lazarides:1985ja}
  G.~Lazarides, C.~Panagiotakopoulos and Q.~Shafi,
  %``BARYOGENESIS AND THE GRAVITINO PROBLEM IN SUPERSTRING MODELS,''
  Phys.\ Rev.\ Lett.\  {\bf 56}, 557 (1986).
  %%CITATION = PRLTA,56,557;%%
  
  %\cite{Lyth:1995hj}
\bibitem{Lyth:1995hj}
  D.~H.~Lyth and E.~D.~Stewart,
  %``Cosmology with a TeV mass GUT Higgs,''
  Phys.\ Rev.\ Lett.\  {\bf 75}, 201 (1995)
  [arXiv:hep-ph/9502417];
  %%CITATION = PRLTA,75,201;%%
  %\cite{Lyth:1995ka}
%\bibitem{Lyth:1995ka}
  %D.~H.~Lyth and E.~D.~Stewart,
  %``Thermal inflation and the moduli problem,''
  Phys.\ Rev.\  D {\bf 53}, 1784 (1996)
  [arXiv:hep-ph/9510204].
  %%CITATION = PHRVA,D53,1784;%%
  
  
  
  %\cite{Berges:2004yj}
\bibitem{Berges:2004yj} 
  J.~Berges,
  %``Introduction to nonequilibrium quantum field theory,''
  AIP Conf.\ Proc.\  {\bf 739}, 3 (2005)
  [hep-ph/0409233].
  %%CITATION = HEP-PH/0409233;%%
  
  %\cite{CalzettaHuBook}
\bibitem{CalzettaHuBook}
  E.~Calzetta and B.~L.~Hu,
  ``Nonequilibrium Quantum Field Theory'',
  Cambridge University Press, Cambridge, UK (2008).
  
  %\cite{BellacBook}
\bibitem{BellacBook}
M. Le Bellac, 
``Thermal field theory,''
Cambridge University Press, Cambridge, UK (2000).

  
  %\cite{Bodeker:2006ij}
\bibitem{Bodeker:2006ij} 
  D.~Bodeker,
  %``Moduli decay in the hot early Universe,''
  JCAP {\bf 0606}, 027 (2006)
  [hep-ph/0605030].
  %%CITATION = HEP-PH/0605030;%%
  
  %\cite{Laine:2010cq}
\bibitem{Laine:2010cq} 
  M.~Laine,
  %``On bulk viscosity and moduli decay,''
  Prog.\ Theor.\ Phys.\ Suppl.\  {\bf 186}, 404 (2010)
  [arXiv:1007.2590 [hep-ph]].
  %%CITATION = ARXIV:1007.2590;%%
  
  %\cite{Arnold:2006fz}
\bibitem{Arnold:2006fz} 
  P.~B.~Arnold, C.~Dogan and G.~D.~Moore,
  %``The Bulk Viscosity of High-Temperature QCD,''
  Phys.\ Rev.\ D {\bf 74}, 085021 (2006)
  [hep-ph/0608012].
  %%CITATION = HEP-PH/0608012;%%
  
  %\cite{Thoma:1994yw}
\bibitem{Thoma:1994yw} 
  M.~H.~Thoma,
  %``Damping of a Yukawa fermion at finite temperature,''
  Z.\ Phys.\ C {\bf 66}, 491 (1995)
  [hep-ph/9406242].
  %%CITATION = HEP-PH/9406242;%%
  
  %\cite{Wang:1999mb}
\bibitem{Wang:1999mb} 
  S.~-Y.~Wang, D.~Boyanovsky, H.~J.~de Vega, D.~S.~Lee and Y.~J.~Ng,
  %``Damping rates and mean free paths of soft fermion collective excitations in a hot fermion gauge scalar theory,''
  Phys.\ Rev.\ D {\bf 61}, 065004 (2000)
  [hep-ph/9902218].
  %%CITATION = HEP-PH/9902218;%%
  
 %\cite{Kiessig:2011fw}
\bibitem{Kiessig:2011fw} 
  C.~Kiessig and M.~Plumacher,
  %``Hard-Thermal-Loop Corrections in Leptogenesis I: CP-Asymmetries,''
  arXiv:1111.1231 [hep-ph].
  %%CITATION = ARXIV:1111.1231;%%
  
  %\cite{Weldon:1982bn}
\bibitem{Weldon:1982bn} 
  H.~A.~Weldon,
  %``Effective Fermion Masses of Order gT in High Temperature Gauge Theories with Exact Chiral Invariance,''
  Phys.\ Rev.\ D {\bf 26}, 2789 (1982).
  %%CITATION = PHRVA,D26,2789;%%
  
  %\cite{Jeon:1994if}
\bibitem{Jeon:1994if} 
  S.~Jeon,
  %``Hydrodynamic transport coefficients in relativistic scalar field theory,''
  Phys.\ Rev.\ D {\bf 52}, 3591 (1995)
  [hep-ph/9409250].
  %%CITATION = HEP-PH/9409250;%%
  
  %\cite{Berera:1998gx}
\bibitem{Berera:1998gx} 
  A.~Berera, M.~Gleiser and R.~O.~Ramos,
  %``Strong dissipative behavior in quantum field theory,''
  Phys.\ Rev.\ D {\bf 58}, 123508 (1998)
  [hep-ph/9803394].
  %%CITATION = HEP-PH/9803394;%%
  
  %\cite{Yokoyama:2005dv}
\bibitem{Yokoyama:2005dv} 
  J.~'i.~Yokoyama,
  %``Can oscillating scalar fields decay into particles with a large thermal mass?,''
  Phys.\ Lett.\ B {\bf 635}, 66 (2006)
  [hep-ph/0510091].
  %%CITATION = HEP-PH/0510091;%%
  
      %\cite{Anisimov:2010dk}
\bibitem{Anisimov:2010dk} 
  A.~Anisimov, W.~Buchmuller, M.~Drewes and S.~Mendizabal,
  %``Quantum Leptogenesis I,''
  Annals Phys.\  {\bf 326}, 1998 (2011)
  [arXiv:1012.5821 [hep-ph]].
  %%CITATION = ARXIV:1012.5821;%%
  
  %\cite{Kubo:1957mj+x}
\bibitem{Kubo:1957mj+x} 
  R.~Kubo,
  %``Statistical mechanical theory of irreversible processes. 1. General theory and simple applications in magnetic and conduction problems,''
  J.\ Phys.\ Soc.\ Jap.\  {\bf 12}, 570 (1957);
  %%CITATION = JUPSA,12,570;%% 
%\cite{Martin:1959jp}
%\bibitem{Martin:1959jp} 
  P.~C.~Martin and J.~S.~Schwinger,
  %``Theory of many particle systems. 1.,''
  Phys.\ Rev.\  {\bf 115}, 1342 (1959).
  %%CITATION = PHRVA,115,1342;%%
 
  %\cite{Dolan:1973qd}
\bibitem{Dolan:1973qd} 
  L.~Dolan and R.~Jackiw,
  %``Symmetry Behavior at Finite Temperature,''
  Phys.\ Rev.\ D {\bf 9}, 3320 (1974).
  %%CITATION = PHRVA,D9,3320;%%
  
  %\cite{Moss:2006gt}
\bibitem{Moss:2006gt} 
  I.~GMoss and C.~Xiong,
  %``Dissipation coefficients for supersymmetric inflatonary models,''
  hep-ph/0603266.
  %%CITATION = HEP-PH/0603266;%%
   
  
    %\cite{Kasuya:1996aq}
\bibitem{Kasuya:1996aq} 
  S.~Kasuya and M.~Kawasaki,
  %``Restriction to parametric resonant decay after inflation,''
  Phys.\ Lett.\ B {\bf 388}, 686 (1996)
  [hep-ph/9603317].
  %%CITATION = HEP-PH/9603317;%%
    
  %\cite{Greene:1998nh}
\bibitem{Greene:1998nh} 
  P.~B.~Greene and L.~Kofman,
  %``Preheating of fermions,''
  Phys.\ Lett.\ B {\bf 448}, 6 (1999)
  [hep-ph/9807339];
  %%CITATION = HEP-PH/9807339;%%
  %\cite{Greene:2000ew}
%\bibitem{Greene:2000ew} 
  %P.~B.~Greene and L.~Kofman,
  %``On the theory of fermionic preheating,''
  Phys.\ Rev.\ D {\bf 62}, 123516 (2000)
  [hep-ph/0003018].
  %%CITATION = HEP-PH/0003018;%%
  
  %\cite{Giudice:1999fb}
\bibitem{Giudice:1999fb} 
  G.~F.~Giudice, M.~Peloso, A.~Riotto and I.~Tkachev,
  %``Production of massive fermions at preheating and leptogenesis,''
  JHEP {\bf 9908}, 014 (1999)
  [hep-ph/9905242];
  %%CITATION = HEP-PH/9905242;%%
  %\cite{Peloso:2000hy}
%\bibitem{Peloso:2000hy} 
  M.~Peloso and L.~Sorbo,
  %``Preheating of massive fermions after inflation: Analytical results,''
  JHEP {\bf 0005}, 016 (2000)
  [hep-ph/0003045].
  %%CITATION = HEP-PH/0003045;%%
  
   %\cite{Allahverdi:2005mz}
\bibitem{Allahverdi:2005mz} 
  R.~Allahverdi and A.~Mazumdar,
  %``Supersymmetric thermalization and quasi-thermal universe: Consequences for gravitinos and leptogenesis,''
  JCAP {\bf 0610}, 008 (2006)
  [hep-ph/0512227].
  %%CITATION = HEP-PH/0512227;%%
  
  %\cite{Allahverdi:2007zz}
\bibitem{Allahverdi:2007zz} 
  R.~Allahverdi and A.~Mazumdar,
  %``Reheating in supersymmetric high scale inflation,''
  Phys.\ Rev.\ D {\bf 76}, 103526 (2007)
  [hep-ph/0603244].
  %%CITATION = HEP-PH/0603244;%%
  
    %\cite{Allahverdi:2011aj}
\bibitem{Allahverdi:2011aj} 
  R.~Allahverdi, A.~Ferrantelli, J.~Garcia-Bellido and A.~Mazumdar,
  %``Non-perturbative production of matter and rapid thermalization after MSSM inflation,''
  Phys.\ Rev.\ D {\bf 83}, 123507 (2011)
  [arXiv:1103.2123 [hep-ph]].
  %%CITATION = ARXIV:1103.2123;%%
  
  %\cite{Allahverdi:2010xz}
\bibitem{Allahverdi:2010xz} 
  R.~Allahverdi, R.~Brandenberger, F.~-Y.~Cyr-Racine and A.~Mazumdar,
  %``Reheating in Inflationary Cosmology: Theory and Applications,''
  Ann.\ Rev.\ Nucl.\ Part.\ Sci.\  {\bf 60}, 27 (2010)
  [arXiv:1001.2600 [hep-th]].
  %%CITATION = ARXIV:1001.2600;%%

  
  %\cite{Kofman:2004yc}
\bibitem{Kofman:2004yc} 
  L.~Kofman, A.~D.~Linde, X.~Liu, A.~Maloney, L.~McAllister and E.~Silverstein,
  %``Beauty is attractive: Moduli trapping at enhanced symmetry points,''
  JHEP {\bf 0405}, 030 (2004)
  [hep-th/0403001].
  %%CITATION = HEP-TH/0403001;%%
  
  
  
  %\cite{Copeland:1995fq}
\bibitem{Copeland:1995fq} 
  E.~J.~Copeland, M.~Gleiser and H.~-R.~Muller,
  %``Oscillons: Resonant configurations during bubble collapse,''
  Phys.\ Rev.\ D {\bf 52}, 1920 (1995)
  [hep-ph/9503217].
  %%CITATION = HEP-PH/9503217;%%
  
  %\cite{McDonald:2001iv}
\bibitem{McDonald:2001iv} 
  J.~McDonald,
  %``Inflaton condensate fragmentation in hybrid inflation models,''
  Phys.\ Rev.\ D {\bf 66}, 043525 (2002)
  [hep-ph/0105235].
  %%CITATION = HEP-PH/0105235;%%
  
  %\cite{Copeland:2002ku}
\bibitem{Copeland:2002ku} 
  E.~J.~Copeland, S.~Pascoli and A.~Rajantie,
  %``Dynamics of tachyonic preheating after hybrid inflation,''
  Phys.\ Rev.\ D {\bf 65}, 103517 (2002)
  [hep-ph/0202031].
  %%CITATION = HEP-PH/0202031;%%
  
  %\cite{Gleiser:2009ys}
\bibitem{Gleiser:2009ys} 
  M.~Gleiser and D.~Sicilia,
  %``A General Theory of Oscillon Dynamics,''
  Phys.\ Rev.\ D {\bf 80}, 125037 (2009)
  [arXiv:0910.5922 [hep-th]].
  %%CITATION = ARXIV:0910.5922;%%
  
  %\cite{Amin:2010dc}
\bibitem{Amin:2010dc} 
  M.~A.~Amin, R.~Easther and H.~Finkel,
  %``Inflaton Fragmentation and Oscillon Formation in Three Dimensions,''
  JCAP {\bf 1012}, 001 (2010)
  [arXiv:1009.2505 [astro-ph.CO]].
  %%CITATION = ARXIV:1009.2505;%%
  
  %\cite{Gleiser:2011xj}
\bibitem{Gleiser:2011xj} 
  M.~Gleiser, N.~Graham and N.~Stamatopoulos,
  %``Generation of Coherent Structures After Cosmic Inflation,''
  Phys.\ Rev.\ D {\bf 83}, 096010 (2011)
  [arXiv:1103.1911 [hep-th]].
  %%CITATION = ARXIV:1103.1911;%%
  
%\cite{Kasuya:2002zs}
\bibitem{Kasuya:2002zs} 
  S.~Kasuya, M.~Kawasaki and F.~Takahashi,
  %``I-balls,''
  Phys.\ Lett.\ B {\bf 559}, 99 (2003)
  [hep-ph/0209358].
  %%CITATION = HEP-PH/0209358;%%
  
  %\cite{Graham:2006xs}
\bibitem{Graham:2006xs} 
  N.~Graham and N.~Stamatopoulos,
  %``Unnatural Oscillon Lifetimes in an Expanding Background,''
  Phys.\ Lett.\ B {\bf 639}, 541 (2006)
  [hep-th/0604134].
  %%CITATION = HEP-TH/0604134;%%
  
  %\cite{Gleiser:2008ty}
\bibitem{Gleiser:2008ty} 
  M.~Gleiser and D.~Sicilia,
  %``Analytical Characterization of Oscillon Energy and Lifetime,''
  Phys.\ Rev.\ Lett.\  {\bf 101}, 011602 (2008)
  [arXiv:0804.0791 [hep-th]].
  %%CITATION = ARXIV:0804.0791;%%
  
  %\cite{Hertzberg:2010yz}
\bibitem{Hertzberg:2010yz} 
  M.~P.~Hertzberg,
  %``Quantum Radiation of Oscillons,''
  Phys.\ Rev.\ D {\bf 82}, 045022 (2010)
  [arXiv:1003.3459 [hep-th]].
  %%CITATION = ARXIV:1003.3459;%%
  
  \bibitem{mukaida}
  K.~Mukaida and K.~Nakayama, in preparation.
  

%%%%%%%%%%%%%%%%%%%%%%%%%%%%%%%%%%%%%%%%%%%%
\end{thebibliography}
\end{document}